\documentclass[a4paper,11pt]{article}
\usepackage{bm,mathrsfs,amsmath,amsthm,amssymb,amscd,longtable,array,graphicx,bm}
\newcommand{\nc}{\newcommand}
\nc{\rnc}{\renewcommand}
\nc{\nn}{\nonumber}
\nc{\der}{{\partial}}
\rnc{\Im}{{\rm{Im}\,}}
\rnc{\Re}{{\rm{Re}\,}}
\nc{\db}{\displaybreak[0]\\}
\nc{\bra}{\langle}
\nc{\ket}{\rangle}
\nc{\bs}{\boldsymbol}

\newtheorem{theorem}{Theorem}[section]

\newtheorem{proposition}[theorem]{Proposition}

\theoremstyle{definition}
\newtheorem{definition}[theorem]{Definition}

\numberwithin{equation}{section}

\numberwithin{equation}{section}

\begin{document}%
%
\title{
A class of partition functions associated with $E_{\tau,\eta}(gl_3)$ \\
by Izergin-Korepin  analysis
}

\author{
Kohei Motegi \thanks{E-mail: kmoteg0@kaiyodai.ac.jp}
\\\\
{\it Faculty of Marine Technology, Tokyo University of Marine Science and Technology,}\\
 {\it Etchujima 2-1-6, Koto-Ku, Tokyo, 135-8533, Japan} \\
\\\\
\\
}

\date{\today}

\maketitle

\begin{abstract}
Recently, a class of partition functions
associated with higher rank rational and trigonometric integrable models
were introduced by  Foda and Manabe.
We use the dynamical $R$-matrix 
of the elliptic quantum group $E_{\tau,\eta}(gl_3)$
to introduce an elliptic analogue of the partition functions associated with $E_{\tau,\eta}(gl_3)$.
We investigate the partition functions of Foda-Manabe type
by developing a nested version of the elliptic Izergin-Korepin analysis,
and present the explicit forms
as symmetrization of multivariable elliptic functions.
We show that special cases are essentially the elliptic weights functions
introduced in the works by
Rim\'anyi-Tarasov-Varchenko, Konno, Felder-Rim\'anyi-Varchenko.
\end{abstract}

\section{Introduction}

Partition functions of integrable models 
\cite{Bethe,Baxter,KBI}
in statistical physics have rich connections
with mathematics and high energy physics.
As for the connection with mathematics,
one of the important facts is that wavefunctions of integrable models
can be expressed using symmetric functions
such as the Schur, Hall-Littlewood, Grothendieck polynomials
and their symplectic analogues, $q$-deformation and
elliptic generalizations.
See
\cite{Bogo,ShigechiUchiyama,BeWh,BWZ,WZnew,vDE,MS,MS2,Korff,GK2,Borodin,Borodin2,BP1,BBF,Iv,BBCG,Tabony,BMN}
for examples on various topics of investigations and applications of
the correspondence between the wavefunctions and symmetric functions.

One of the challenging problems is
to go beyond six-vertex models 
and study partition functions for higher rank models.
See \cite{TarVarSigma,FWZ,FW,WZ,Tak,BW,metaplectic,BBBG,BBBG2,BSW}
for examples on seminal and recent works on this topic.
One of the recent progresses has been made by Foda and Manabe \cite{FM},
which they introduced a new class of partition functions
for higher rank rational and trigonometric models,
motivated by the Bethe/Gauge correspondence \cite{NS1,NS2},
and their partition functions seem to deserve further studies.

In this paper, we introduce and study an elliptic analogue of
the partition functions of Foda-Manabe type associated with elliptic quantum group
\cite{Felder,FV1,FV2,Ca} $E_{\tau,\eta}(gl_3)$.
We use the Izergin-Korepin method for the analysis of the partition functions.
The Izergin-Korepin method is a method initiated by Korepin \cite{Ko} and used by Izergin \cite{Iz} to find a determinant representation for the domain wall boundary partition functions of the trigonometric six-vertex model.
The determinant representation (Izegin-Korepin determinant)
has found many applications and connections to
many branches of mathematics and mathematical physics,
such as the enumeration of the alternating sign matrices,
relations with orthogonal polynomials and classical integrable systems
\cite{Br,Ku1,Ku2,Okada,CP,KZ,BL,HK1,HK2}.
The Izergin-Korepin method was also applied
to variants of the domain wall boundary partition functions
and extended to the scalar products \cite{Ku1,Ku2,Tsuchiya,Wheeler}.
There are also developments on the studies of the
domain wall boundary partition functions
for elliptic integrable models by various methods.
See
\cite{FWZ,PRS,Ros,FK,Chinesegroup,Chinesegroup2,Galleasone,Galleasthree,GL,Lamers} for examples on this topic.

Recently, for the case of six-vertex type models,
the Izergin-Korepin method was extended to the wavefunctions
\cite{Motegi,Motegi2},
and we develop a nested version of this method in this paper
for the purpose of analyzing the partition functions of Foda-Manabe type.
We use the Izergin-Korepin method in two steps.
We first use the method to analyze partition functions
which we call as the base partition functions, which is essentially partition functions
associated with $E_{\tau,\eta}(gl_2)$.
We next perform the Izergin-Korepin analysis on the partition functions of Foda-Manabe type
associated with $E_{\tau,\eta}(gl_3)$.
The Izergin-Korepin method is a method which uses graphical representations
of integrable models to construct relations between partition functions of different sizes.
One needs the intitial condition, and the base partition functions
essentially serve as the initial condition for the second Izergin-Korepin analysis.
This is similar to the Izergin-Korepin analysis on the scalar products by Wheeler \cite{Wheeler},
which he introduced intermediate scalar products as a generalization, and the initial condition
of the Izergin-Korepin analysis for the intermediate scalar products is essentially given by the domain wall boundary partition functions.
Foda and Manabe mention that the partition functions of rational and trigonometric models they introduced contain the nested wavefunctions as special cases.
From the point of view of the Izergin-Korepin analysis, the relation between
the partition functions they introduced and the nested wavefunctions
resemble the relation between the intermediate scalar products and the scalar products,
since one needs generalizations of the partition functions to investigate the
original ones.
We also show that special cases are essentially
the elliptic weights functions introduced in the works by
Rim\'anyi-Tarasov-Varchenko, Konno, Felder-Rim\'anyi-Varchenko
\cite{Konno1,Konno2,FRV,RTV2},
which appear as objects in the integral representation of
the solutions to the elliptic $q$-KZ equations, and
also essentially play the role of elliptic stable envelope maps
for the cotangent bundles of flag varieties,
which are geometric objects
originally proposed by Aganagic-Okounkov \cite{AO}
as an elliptic generalization of the 
cohomological stable envelopes
which appear in the works by Maulik-Okounkov \cite{MO},
in which they
initiated a program to relate quantum torus equivariant cohomology
of quiver varieties and representation theory of quantum groups,
which is considered as a mathematical formulation of
the Bethe/gauge correspondence \cite{NS1,NS2}.

This paper is organized as follows.
In the next section, we introduce the dynamical $R$-matrix and list the
properties of theta functions which are necessary for the present paper.
In section 3, we introduce two types of partition functions:
the base partition functions and partition functions of Foda-Manabe type.
In section 4, we analyze the base partition functions.
In section 5, we perform the Izergin-Korepin analysis on
partition functions of Foda-Manabe type, and determine
the explicit form of the partition functions.
In section 6, we show that special cases of the elliptic partition functions
of Foda-Manabe type are the elliptic weights functions studied in
Rim\'anyi-Tarasov-Varchenko, Konno, Felder-Rim\'anyi-Varchenko.
Section 7 is devoted to the conclusion of this paper.

\section{Theta function and Dynamical $R$-matrix}

In this section, we recall the properties of the theta functions 
and the dynamical $R$-matrix
which we use in this paper.

First we introduce the theta function
\begin{align}
[z]=-\sum_{j \in {\bf Z}+1/2}e^{i \pi j^2 \tau+2 \pi i j (z+1/2)},
\label{theta}
\end{align}
which is an odd function $[-z]=-[z]$
and satisfy the quasi-periodicities
\begin{align}
[z+1]&=-[z], \\
[z+\tau]&=-e^{-2 \pi i z-\pi i \tau} [z].
\end{align}

For the elliptic version of the Izergin-Korepin analysis,
we use the following facts about the elliptic polynomials \cite{PRS,FSfelderhof}.

A character is a group homomorphism
$\chi$ from multiplicative groups
$\Gamma=\mathbf{Z}+\tau \mathbf{Z}$ to $\mathbf{C}^\times$.
An $n$-dimensional space $\Theta_n(\chi)$
is defined for each character $\chi$ and positive integer $n$,
which consists of holomorphic functions $\phi(y)$ on $\mathbf{C}$
satisfying the quasi-periodicities
\begin{align}
\phi(y+1)&=\chi(1) \phi(y), \label{propertyuseone} \\
\phi(y+\tau)&=\chi(\tau) e^{-2 \pi i ny-\pi i n \tau}\phi(y).
\label{propertyusetwo}
\end{align}
The elements of the space $\Theta_n(\chi)$ are called elliptic polynomials.
The space $\Theta_n(\chi)$ is $n$-dimensional \cite{PRS,FSfelderhof}
and the following fact holds for the elliptic polynomials.
\begin{proposition} \cite{PRS,FSfelderhof} \label{propositionelliptic}
Suppose there are two elliptic polynomials $P(y)$ and $Q(y)$
in $\Theta_n(\chi)$, where $\chi(1)=(-1)^n$, $\chi(\tau)=(-1)^n e^\alpha$.
If those two polynomials are equal $P(y_j)=Q(y_j)$
at $n$ points $y_j$, $j=1,\dots,n$ satisfying
$y_j-y_k \not\in \Gamma$, $\sum_{k=1}^N y_k-\alpha \not\in \Gamma$,
then the two polynomials are exactly the same $P(y)=Q(y)$.
\end{proposition}

The above proposition is an elliptic analogue of the
following properties for ordinary polynomials:
if $P(y)$ and $Q(y)$ are polynomials of degree $n-1$ in $y$,
and if these polynomials match at $n$ distinct points,
then the two polynomials are exactly the same.
These properties ensure the uniqueness of the Izergin-Korepin analysis,
which was effectively used in \cite{PRS} to study
the domain wall boundary partition functions of
the Andrews-Baxter-Forrester model \cite{ABF}
which is related to the eight-vertex model \cite{eightvertex}
by the vertex-face transformation. \\
\\

Next, let us reall the dynamical $R$-matrix.
We use the dynamical $R$-matrix for the face-type
elliptic quantum group $E_{\tau,\eta}(gl_n)$ \cite{Felder,FV1,FV2,Ca}
(there are also vertex-type and centrally-extended versions
of the elliptic quantum groups
\cite{FIJKMY,Fr,Konno,JKOS}).
The dynamical $R$-matrix for the elliptic quantum group
$E_{\tau,\eta}(gl_n)$ is a function
$R(z,\lambda): {\bf C} \otimes \mathfrak{h}^*
\longrightarrow \mathrm{End}(V \otimes V)$
where $\mathfrak{h}$ is a Cartan subalgebra of $gl_n$,
$\mathfrak{h}^*$ is its dual
and $V$  is a diagonalizable $\mathfrak{h}$-module.
For the case we consider,
$V$ is the $\mathfrak{h}$-module ${\bf C}^n$
with standard basis $e_i, i=1,\dots,n$.
Let us define $\mu_i \in \mathfrak{h}^*$ as
$\mu_i(h)=h^i$ if $h=\mathrm{diag}(h^1,\dots,h^n) \in \mathfrak{h}$.
Then $V=\oplus_{i=1}^N V^{\mu_i}$ where $V^{\mu_i}={\bf C}e_i$.

The explicit form of the dynamical $R$-matrix is given by
\begin{align}
R(z,\lambda)=\sum_{i=1}^N [z-\gamma] E_{ii} \otimes E_{ii}
+\sum_{i \neq j} \alpha(z,\lambda_i-\lambda_j) E_{ii} \otimes E_{jj}
+\sum_{i \neq j} \beta(z,\lambda_i-\lambda_j)E_{ij} \otimes E_{ji},
\label{dynamicalrmatrix}
\end{align}
where $E_{ij}$ are matrix units $E_{ij}e_k=\delta_{jk}e_i$.
$\alpha(z,\lambda)$ and $\beta(z,\lambda)$ are given
in terms of theta functions as
\begin{align}
\alpha(z,\lambda)=\frac{[z][\lambda+\gamma]}{[\lambda]},
\ \beta(z,\lambda)=-\frac{[z+\lambda][\gamma]}{[\lambda]},
\end{align}
and $\lambda_i$ in \eqref{dynamicalrmatrix} are coordinate functions
$\lambda_i=\lambda(E_{ii}), i=1,\dots,n$.

\begin{figure}[ht]
\includegraphics[width=15cm]{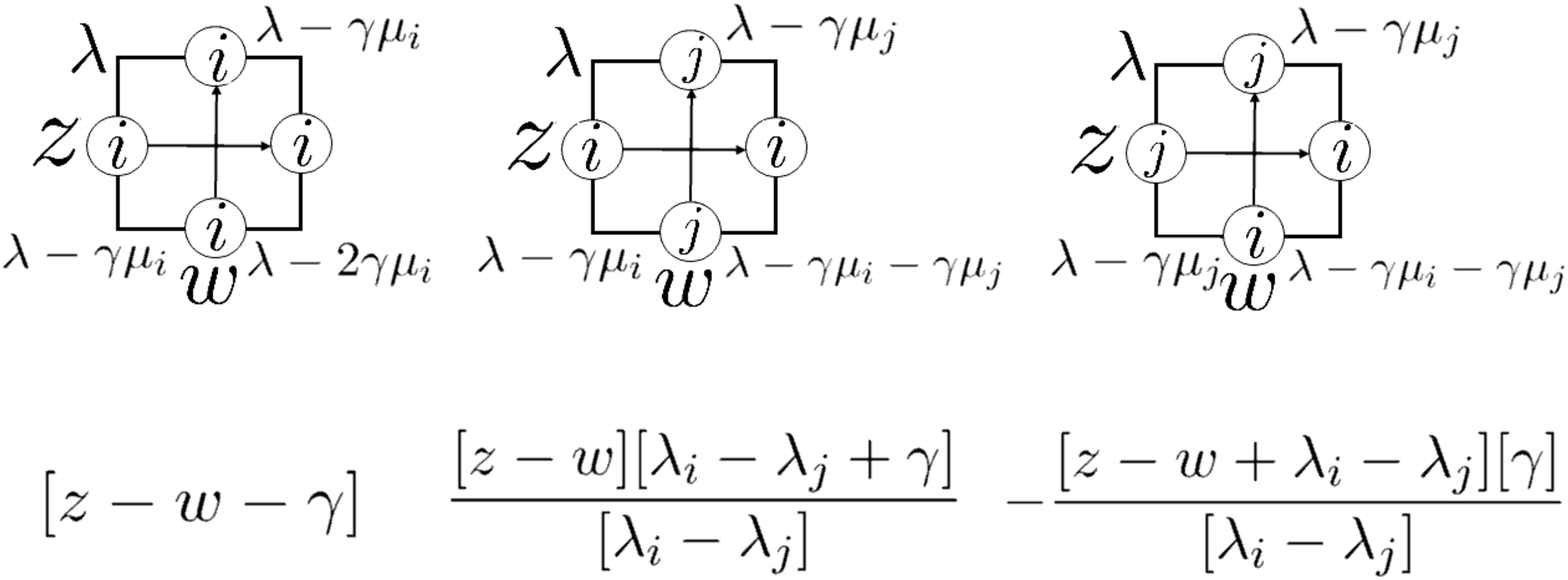}
\caption{A graphical description of the dynamical $R$-matrix $R(z-w,\lambda)$
\eqref{dynamicalrmatrix}. Here, $i$ and $j$ are different $i \neq j$.
}
\label{picturedynamicalrmatrix}
\end{figure}

The dynamical $R$-matrix \eqref{dynamicalrmatrix} satisfies the
dynamical Yang-Baxter relation
\begin{align}
&R_{23}(z_2-z_3,\lambda)
R_{13}(z_1-z_3,\lambda-\gamma h_2)
R_{12}(z_1-z_2,\lambda) \nonumber \\
=&R_{12}(z_1-z_2,\lambda-\gamma h_3)
R_{13}(z_1-z_3,\lambda)
R_{23}(z_1-z_3,\lambda-\gamma h_1),
\end{align}
acting on $V_1 \otimes V_2 \otimes V_3$.
The subscripts indicate the spaces the operators are acting on.
For example,
\begin{align}
R_{12}(z_1-z_2,\lambda-\gamma h_3)(v_1 \otimes v_2 \otimes v_3)
=R(z_1-z_2,\lambda-\gamma \alpha)v_1 \otimes v_2 \otimes v_3,
\end{align}
if $v_3 \in V^\alpha$.

The dynamical $R$-matrix has its origin
in the elliptic face model \cite{ABF,DJKMO,JKMO},
and it describes the face model like a six-vertex model with an
additional dynamical parameter.
The dynamical $R$-matrix $R(z-w,\lambda)$ can be expressed
as Figure \ref{picturedynamicalrmatrix} for example.
We use this graphical description of the dynamical $R$-matrix
to construct and study partition functions.

\section{Partition functions}

We introduce two types of partition functions in this section:
the base partition functions and
an elliptic analogue of the
partition functions introduced by Foda and Manabe \cite{FM} recently.

First, we introduce monodromy matrix as
\begin{align}
&T_a(z|w_1,\dots,w_L|\lambda) \nonumber \\
=&R_{aL}(z-w_{L},\lambda-\gamma (h_1+\cdots+h_{L-1})) \cdots
R_{a2}(z-w_2,\lambda-\gamma h_1) R_{a1}(z-w_1,\lambda),
\end{align}
acting on $V_a \otimes (V_1 \otimes \cdots \otimes V_L)$.
We also use bra-ket notations.
We denote the basis vector $e_i$ on $V_j$
as $| i \rangle_j$
and its dual $e_i^*$ as ${}_j \langle i|$.

\begin{figure}[ht]
\includegraphics[width=12cm]{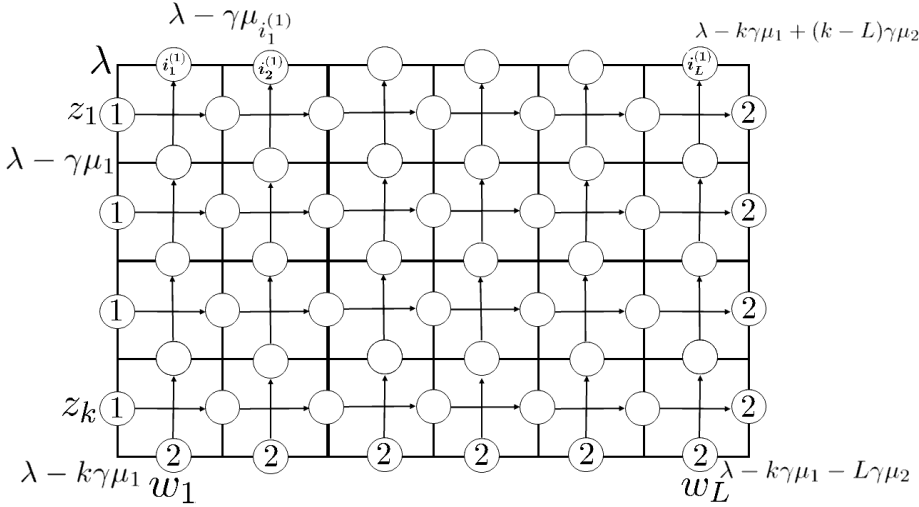}
\caption{The base partition functions
$W_{L,k}(z_1,\dots,z_k|w_1,\dots,w_L|I_1,\dots,I_k|\lambda)$
\eqref{basewavefunctions}.
}
\label{picturebasepartitionfunction}
\end{figure}

We now introduce the following partition functions
(Figure \ref{picturebasepartitionfunction})

\begin{align}
&W_{L,k}(z_1,\dots,z_k|w_1,\dots,w_L|I_1,\dots,I_k|\lambda)
\nonumber \\
=&{}_{a_1} \langle 2| \otimes \cdots \otimes
{}_{a_k} \langle 2| \otimes
{}_1 \langle i_1^{(1)} | \otimes
\cdots \otimes {}_L \langle i_{L}^{(1)} |
\nonumber \\
&T_{a_1}(z_1|w_1,\dots,w_L|\lambda)
T_{a_2}(z_2|w_1,\dots,w_L|\lambda-\gamma h_{a_1}) \cdots
\nonumber \\
&\cdots
T_{a_k}(z_k|w_1,\dots,w_L|\lambda-\gamma (h_{a_1}+\cdots h_{a_{k-1}}))
|1 \rangle_{a_1} \otimes \cdots \otimes |1 \rangle_{a_k}
\otimes
|2 \rangle_1 \otimes \cdots \otimes |2 \rangle_L,
\label{basewavefunctions}
\end{align}
where $i_j^{(1)}$ ($j=1,\dots, L$) is 1 if $j=I_\ell$ for some $\ell \in \{1,\dots,k \}$, and 2 otherwise.
In this paper, we call the partition functions
$W_{L,k}(z_1,\dots,z_k|w_1,\dots,w_L|I_1,\dots,I_k|\lambda)$
as the base partition functions.
Note that only matrix elements of the form
${}_{1} \langle j_1 | {}_{2} \langle j_2| R_{12}(z-w,\lambda)
| i_1 \rangle_1  | i_2 \rangle_2$, $i_1,i_2,j_1,j_2=1,2$
contribute to the base partition functions, i.e.
the base partition functions are essentially partition functions
associated with $E_{\tau,\eta}(gl_2)$.

Next we introduce a class of
partition functions associated with $E_{\tau.\eta}(gl_3)$,
which is an elliptic analogue of the one
recently introduced by Foda and Manabe \cite{FM}.
Let us denote
\begin{align}
&T_a(z|\{\bm z^{(2)} \}, \{\bm w^{(1)} \}|\lambda) \nonumber \\
=&R_{a, k_2+L_1}(z-w_{L_1}^{(1)},\lambda-\gamma (h_1+\cdots+h_{k_2+L_1-1})) \cdots
R_{a, k_2+1}(z-w_{1}^{(1)},\lambda-\gamma (h_1+\cdots+h_{k_2}))
\nonumber \\
\times&R_{ak_2}(z-z_{k_2}^{(2)},\lambda-\gamma (h_1+\cdots+h_{k_2-1}))
\cdots
R_{a1}(z-z_1^{(2)},\lambda),
\end{align}

\begin{figure}[ht]
\includegraphics[width=15cm]{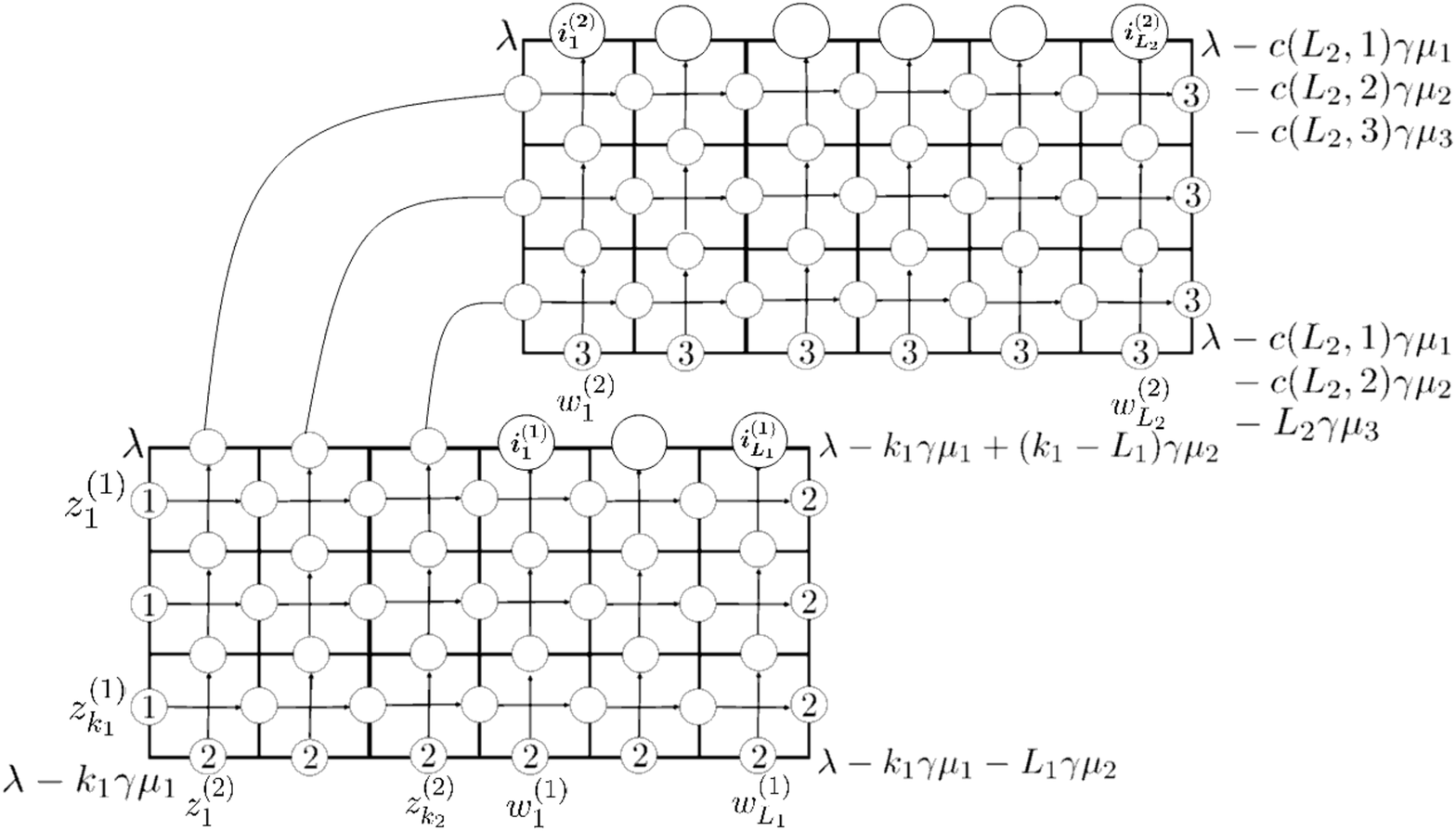}
\caption{The partition functions of Foda-Manabe type
associated with $E_{\tau,\eta}(gl_3)$
$W(\{ {\bm z^{(1)}} \}, \{ {\bm z^{(2)}} \}
|\{ {\bm w^{(1)}} \}, \{ {\bm w^{(2)}} \}
|\{k_1,k_2,L_1,L_2,{\bm I} \}|\lambda)$
\eqref{fodamanabepartitionfunctions}.
}
\label{picturehigherrankpartitionfunction}
\end{figure}

The partition functions we consider is (Figure \ref{picturehigherrankpartitionfunction})

\begin{align}
&W(\{ {\bm z^{(1)}} \}, \{ {\bm z^{(2)}} \}
|\{ {\bm w^{(1)}} \}, \{ {\bm w^{(2)}} \}|\{k_1,k_2,L_1,L_2,{\bm I} \}|\lambda)
\nonumber \\
=&\sum_{\ell_{1},\dots,\ell_{k_2}=1,2}
{}_{a_1} \langle 3| \otimes \cdots \otimes
{}_{a_{k_2}} \langle 3| \otimes
{}_{1} \langle i_1^{(2)} | \otimes
\cdots \otimes {}_{L_2} \langle i_{L_2}^{(2)} |
\nonumber \\
&T_{a_1}(z_1^{(2)}|w_1^{(2)},\dots,w_{L_2}^{(2)}|\lambda)
T_{a_2}(z_2^{(2)}|w_1^{(2)},\dots,w_{L_2}^{(2)}|\lambda-\gamma h_{a_1}) \cdots
\nonumber \\
&\cdots
T_{a_{k_2}}(z_{k_2}^{(2)}|w_1^{(2)},\dots,w_{L_2}^{(2)}|\lambda-\gamma (h_{a_1}+\cdots+ h_{a_{{k_2}-1}}))
|\ell_1 \rangle_{a_1} \otimes \cdots \otimes |\ell_{k_2} \rangle_{a_{k_2}}
\otimes
|3 \rangle_1 \otimes \cdots \otimes |3 \rangle_{L_2}
\nonumber \\
\times&{}_{a_1} \langle 2| \otimes \cdots \otimes
{}_{a_{k_1}} \langle 2| \otimes
{}_{1} \langle \ell_{1} | \otimes
\cdots \otimes {}_{k_2} \langle \ell_{k_2} | \otimes
{}_{k_2+1} \langle i_1^{(1)} | \otimes
\cdots \otimes {}_{k_2+L_1} \langle i_{L_1}^{(1)} |
\nonumber \\
&T_{a_1}(z_1^{(1)}|\{{\bm z^{(2)}} \}, \{{\bm w^{(1)}} \}|\lambda)
T_{a_2}(z_2^{(1)}|\{{\bm z^{(2)}} \}, \{{\bm w^{(1)}} \}|\lambda-\gamma h_{a_1}) \cdots
\nonumber \\
&\cdots
T_{a_{k_1}}(z_{k_1}^{(1)}|\{{\bm z^{(2)}} \}, \{{\bm w^{(1)}} \}|\lambda-\gamma (h_{a_1}+\cdots+ h_{a_{{k_1}-1}}))
|1 \rangle_{a_1} \otimes \cdots \otimes |1 \rangle_{a_{k_1}}
\otimes
|2 \rangle_1 \otimes \cdots \otimes |2 \rangle_{k_2+L_1}.
\label{fodamanabepartitionfunctions}
\end{align}

The partition functions
$
W(\{ {\bm z^{(1)}} \}, \{ {\bm z^{(2)}} \}
|\{ {\bm w^{(1)}} \}, \{ {\bm w^{(2)}} \}
|\{k_1,k_2,L_1,L_2,{\bm I} \}|\lambda)$ depend on
the sets of parameters
$\{ {\bm z^{(1)}} \}=\{z_1^{(1)},\dots,z_{k_1}^{(1)} \}$, 
$\{ {\bm z^{(2)}} \}=\{z_1^{(1)},\dots,z_{k_2}^{(2)} \}$,
$\{ {\bm w^{(1)}} \}=\{w_1^{(1)},\dots,w_{L_1}^{(1)} \}$
$\{ {\bm w^{(2)}} \}=\{w_1^{(2)},\dots,w_{L_2}^{(2)} \}$.
The partition functions of Foda-Manabe type also depend on
``the configuration of colors''
$i_1^{(1)},\dots,\dots,i_{L_1}^{(1)} \in \{1,2 \}$,
$i_1^{(2)},\dots,\dots,i_{L_2}^{(2)} \in \{1,2,3 \}$.

We adopt the label introduced by Foda-Manabe
to label the configurations.
A configuration of colors is labeled by a
set 
${\bm I}:=\{{\bm I_{k_1}^{(1)}}, {\bm I_{k_2}^{(2)}},
 \widehat{{\bm I}}_{k_2+L_1}^{(2)}, \widehat{{\bm I}}_{L_2}^{(3)} \}$
where the subsets
${\bm I_{k_1}^{(1)}}, {\bm I_{k_2}^{(2)}},
 \widehat{{\bm I}}_{k_2+L_1}^{(2)}, \widehat{{\bm I}}_{L_2}^{(3)} $
satisfy the following relations
\begin{align}
&{\bm I}_{k_1}^{(1)} \subset \widehat{{\bm I}}_{k_2+L_1}^{(2)}
:={\bm I_{k_2}^{(2)}} \cup \{ L_2+1,\dots,L_2+L_1 \}, \\
&{\bm I_{k_2}^{(2)}} \subset \widehat{{\bm I}}_{L_2}^{(3)}:=\{1,\dots,L_2 \}.
\end{align}
The relation between the naive label
$i_1^{(1)},\dots,\dots,i_{L_1}^{(1)}$,
$i_1^{(2)},\dots,\dots,i_{L_2}^{(2)}$
for the configuration of colors
and ${\bm I}=\{{\bm I_{k_1}^{(1)}}, {\bm I_{k_2}^{(2)}},
\widehat{{\bm I}}_{k_2+L_1}^{(2)}, \widehat{{\bm I}}_{L_2}^{(3)} \}$
is as follows.
We label the positions which have colors $i_1^{(2)},\dots,\dots,i_{L_2}^{(2)}$
as $1,2,\dots,L_2$,
and positions which have colors $i_1^{(1)},\dots,\dots,i_{L_1}^{(1)}$
are labeled as $L_2+1,\dots,L_1+L_2$.
${\bm I}_{k_2}^{(2)}$ is a set such that
$\widehat{{\bm I}}_{L_2}^{(3)} \backslash {\bm I_{k_2}^{(2)}}$
is the set of positions which have color $3$,
and ${\bm I}_{k_1}^{(1)}$ is the set of positions which have color $1$
so that $\widehat{{\bm I}}_{k_2+L_1}^{(2)} \backslash {\bm I}_{k_1}^{(1)}$
becomes the set which records the positions of color $2$.
Since the definition of the subsets of ${\bm I}$
depends on integers $k_1,k_2,L_1,L_2$ which 
characterize the sizes of partition functions,
we introduce the full label $\{k_1,k_2,L_1,L_2,{\bm I} \}$
for the configuration of colors.
This full label is important for the Izergin-Korepin analysis
for the partition functions of Foda-Manabe type.
In this paper, we call the quadruplet $\{k_1,k_2,L_1,L_2 \}$ as the size of
the partition functions.
Also note that
throughout this paper, when one denotes the set as
${\bm I}_{k}^{(b)}$, the number of elements of the set
is $k$, and the elements of the set
are denoted as $I_{1}^{(b)}, \cdots, I_k^{(b)}$ where
$I_{1}^{(b)} < \cdots < I_k^{(b)}$.

For later purpose, we also use
the induced set $\widetilde{{\bm I_{k_1}^{(1)}}}$
which is induced by the map
\begin{align}
{\bm I}_{k_1}^{(1)} \subset \widehat{{\bm I}}_{k_2+L_1}^{(2)}
\longrightarrow \widetilde{{\bm I_{k_1}^{(1)}}} \subset
\{1,\dots,k_2+L_1 \}. \label{maptoinduce}
\end{align}
We map the set
$\widehat{{\bm I}}_{k_2+L_1}^{(2)}$
to $\{1,\dots,k_2+L_1 \}$ by $\widehat{I}_{a}^{(2)} \longrightarrow a$
$(a=1,\dots,k_2+L_1)$,
and correspondingly the elements in
${\bm I}_{k_1}^{(1)}$ which are included in $\widehat{{\bm I}}_{k_2+L_1}^{(2)}$
are naturally mapped to elements in $\{1,\dots,k_2+L_1 \}$,
which form the induced subset $\widetilde{{\bm I_{k_1}^{(1)}}}$.

\section{Base partition functions}
In this section,
we analyze the base partition functions
$W_{L,k}(z_1,\dots,z_k|w_1,\dots,w_L|I_1,\dots,I_k|\lambda)$
which is used as the initial condition for the
Izergin-Korepin analysis \cite{Ko,Iz} on the partition functions
of Foda-Manabe type
$W(\{ {\bm z^{(1)}} \}, \{ {\bm z^{(2)}} \}
|\{ {\bm w^{(1)}} \}, \{ {\bm w^{(2)}} \}
|\{k_1,k_2,L_1,L_2,{\bm I} \}|\lambda)$ in the next section.
The idea of the Izergin-Korepin analysis
is to construct relations between partition functions
of different sizes and determine the initial condition,
and find the explicit forms satisfying the recursive relations
and the initial condition.
The base partition functions
$W_{L,k}(z_1,\dots,z_k|w_1,\dots,w_L|I_1,\dots,I_k|\lambda)$
serve as the initial condition for the Izergin-Korepin analysis on
the partition functions
$W(\{ {\bm z^{(1)}} \}, \{ {\bm z^{(2)}} \}
|\{ {\bm w^{(1)}} \}, \{ {\bm w^{(2)}} \}
|\{k_1,k_2,L_1,L_2,{\bm I} \}|\lambda)$.
In this section, we analyze the base partition functions
$W_{L,k}(z_1,\dots,z_k|w_1,\dots,w_L|I_1,\dots,I_k|\lambda)$
itself by using the Izergin-Korepin method for the wavefunctions
\cite{Motegi,Motegi2}.
First, we introduce the elliptic multivariable functions
and state the correspondence with
the base partition functions.

\begin{definition}
We define symmetric functions
$E_{L,k}(z_1,\dots,z_k|w_1,\dots,w_L|I_1,\dots,I_k|\lambda)$
that depend on the symmetric variables $z_1,\dots,z_k$,
complex parameters $w_1,\dots,w_L$, $\gamma$, $\lambda_1, \lambda_2$
and integers $I_1,\dots,I_k$ satisfying
$1 \le I_1 < \cdots < I_k \le L$,
\begin{align}
&E_{L,k}(z_1,\dots,z_k|w_1,\dots,w_L|I_1,\dots,I_k|\lambda)
\nonumber \\
=&\frac{[\gamma]^k}{\prod_{j=1}^k [\lambda_1-\lambda_2+(1-j)\gamma]}
\sum_{\sigma \in S_k} 
\prod_{a=1}^k
\Bigg(
[z_{\sigma(a)}-w_{I_a}+\lambda_2-\lambda_1+(2a-1-I_a)\gamma]
\nonumber \\
\times&\prod_{i=1}^{I_a-1}[z_{\sigma(a)}-w_i]
\prod_{i=I_a+1}^L[z_{\sigma(a)}-w_i-\gamma]
\Bigg)
\prod_{1 \le a < b \le k}
\frac{[z_{\sigma(a)}-z_{\sigma(b)}+\gamma]}{[z_{\sigma(a)}-z_{\sigma(b)}]}.
\label{basesymmetricfunction}
\end{align}
\end{definition}

\begin{theorem} \label{basetheorem}
The base partition functions
$W_{L,k}(z_1,\dots,z_k|w_1,\dots,w_L|I_1,\dots,I_k|\lambda)$
can be explicitly expressed as the
symmetric functions
$E_{L,k}(z_1,\dots,z_k|w_1,\dots,w_L|I_1,\dots,I_k|\lambda)$,
\begin{align}
W_{L,k}(z_1,\dots,z_k|w_1,\dots,w_L|I_1,\dots,I_k|\lambda)
=E_{L,k}(z_1,\dots,z_k|w_1,\dots,w_L|I_1,\dots,I_k|\lambda).
\end{align}
\end{theorem}

\begin{figure}[ht]
\includegraphics[width=12cm]{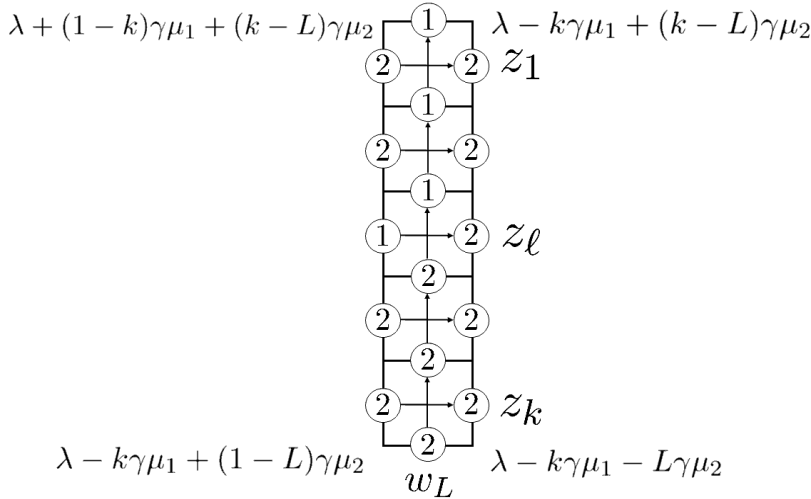}
\caption{The states of the rightmost column of the partition functions
$W_{L,k}(z_1,\dots,z_k|w_1,\dots,w_L|I_1,\dots,I_k|\lambda)$
giving non-zero contributions are graphically represented
as above.
The factors including $w_L$ all come from
the matrix elements of the dynamical $R$-matrices in
this column and can be computed from the above graphical description.
We get
$\displaystyle g_\ell(w_L)=
[z_\ell-w_L+\lambda_2-\lambda_1+(2k-L-\ell) \gamma]
\prod_{j=1}^{\ell-1} [z_j-w_L]
\prod_{j=\ell+1}^k [z_j-w_L-\gamma], \ \ \ \ell=1,\cdots,k$.
}
\label{pictureforqpbase}
\end{figure}

\begin{figure}[ht]
\includegraphics[width=12cm]{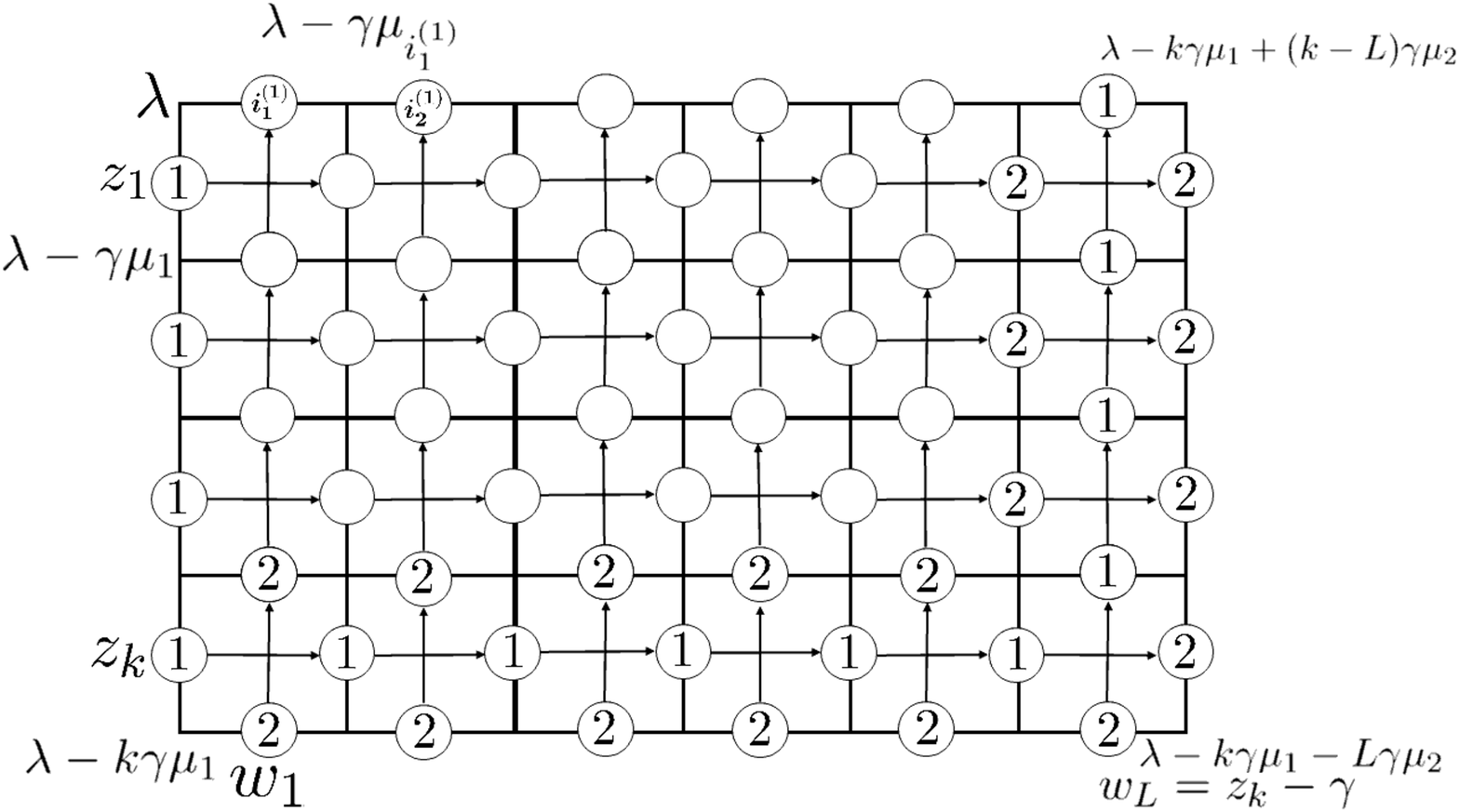}
\caption{The base partition functions satisfying $I_k=L$,
evaluated at $w_L=z_k-\gamma$ \eqref{baserecursionwavefunction}.
The dynamical $R$-matrices
at the rightmost column and the bottom row are all frozen.
}
\label{picturebasepartitionfunctionrecursion}
\end{figure}

\begin{figure}[ht]
\includegraphics[width=12cm]{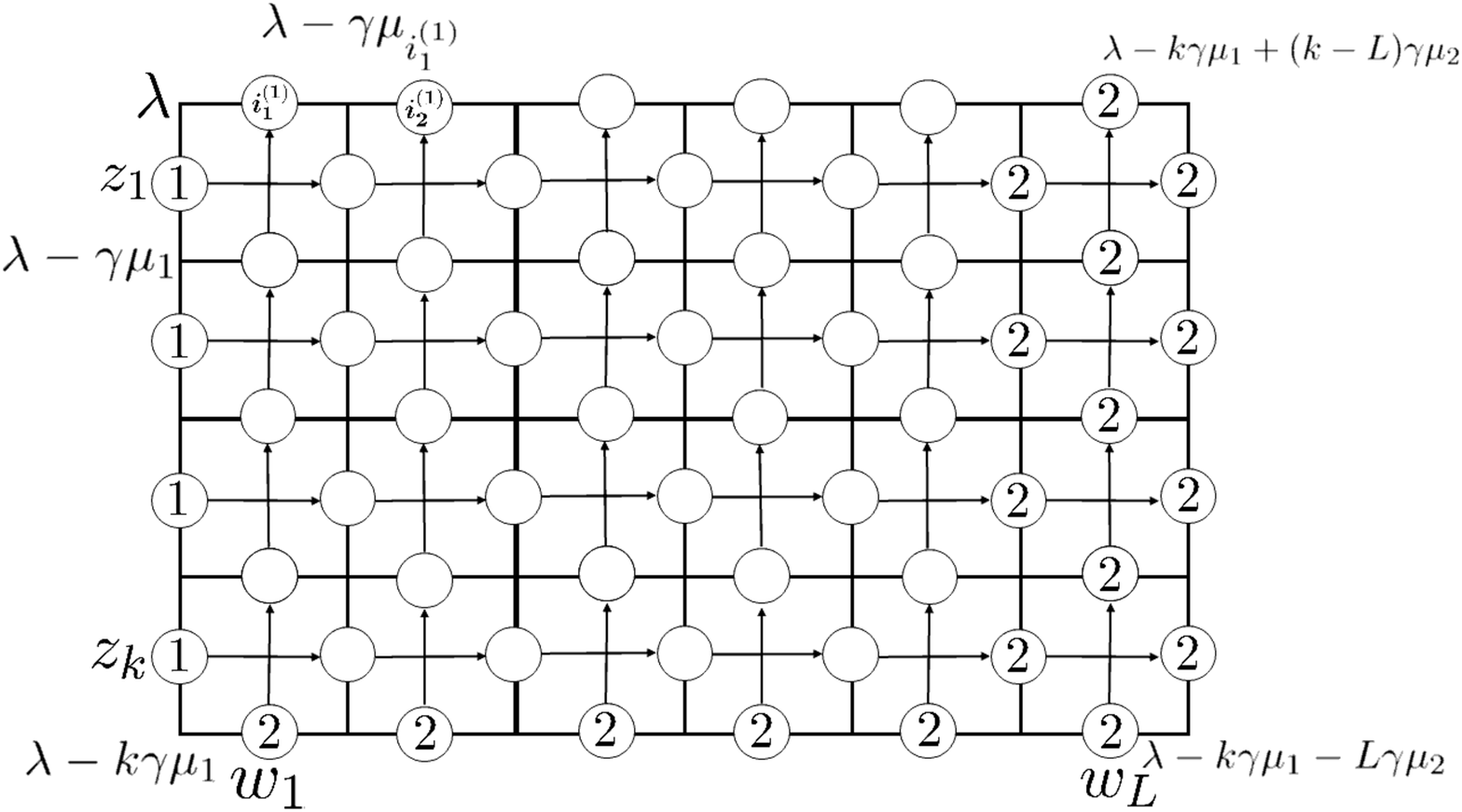}
\caption{The base partition functions with $I_k \neq L$
\eqref{baserecursionwavefunction2}. The dynamical $R$-matrices at the rightmost column are all frozen.
}
\label{picturebasepartitionfunctionfactorization}
\end{figure}
We give below
a proof of Theorem \ref{basetheorem} by using
the Izergin-Korepin method \cite{Ko,Iz}
for the wavefunctions \cite{Motegi,Motegi2}.
See also \cite{Borodin,FV2} which treat the same
type of elliptic partition functions by different methods.

\begin{proof}
First, we construct Korepin's lemma, i.e. list the properties
of the base partition functions which uniquely define them.
For the case of partition functions of wavefunctions type,
it is given by the following proposition.
\begin{proposition}
\label{basewavefunctionizerginkorepin}
The base partition functions
$W_{L,k}(z_1,\dots,z_k|w_1,\dots,w_L|I_1,\dots,I_k|\lambda)$
possess the following properties. \\
\\
 (1) If $I_k=L$, the base partition functions
$W_{L,k}(z_1,\dots,z_k|w_1,\dots,w_L|I_1,\dots,I_k|\lambda)$
are elliptic polynomials of $w_L$ in $\Theta_k(\chi)$ with quasi-periodicities
\begin{align}
&W_{L,k}(z_1,\dots,z_k|w_1,\dots,w_L+1|I_1,\dots,I_k|\lambda)
\nonumber \\
=&(-1)^k W_{L,k}(z_1,\dots,z_k|w_1,\dots,w_L|I_1,\dots,I_k|\lambda),
\label{qpbaseone}
\\
&W_{L,k}(z_1,\dots,z_k|w_1,\dots,w_L+\tau|I_1,\dots,I_k|\lambda),
\nonumber \\
=&(-1)^k \mathrm{exp}
\Bigg(-2 \pi i 
\Bigg(
k w_L-\sum_{i=1}^k z_i+\lambda_1-\lambda_2+\gamma (L-k)
\Bigg)
- \pi i k \tau \Bigg) \nonumber \\
&\times W_{L,k}(z_1,\dots,z_k|w_1,\dots,w_L|I_1,\dots,I_k|\lambda).
\label{qpbasetwo}
\end{align}
\\
 (2) The base partition functions 
$W_{L,k}(z_1,\dots,z_k|w_1,\dots,w_L|I_1,\dots,I_k|\lambda)$
are symmetric with respect to $z_1,\dots,z_k$.
\\
\\
(3) If $I_k=L$, the following recursive relations between the
base partition functions hold (Figure \ref{picturebasepartitionfunctionrecursion}):
\begin{align}
&W_{L,k}(z_1,\dots,z_k|w_1,\dots,w_L|I_1,\dots,I_k|\lambda)
|_{w_L=z_k-\gamma}
\nonumber \\
=&\frac{[\gamma][\lambda_2-\lambda_1+(2k-L)\gamma]}{[\lambda_1-\lambda_2+(1-k)\gamma]}
\prod_{j=1}^{k-1} [z_j-z_k+\gamma]
\prod_{j=1}^{L-1} [z_k-w_j]
\nonumber \\
&\times W_{L-1,k-1}(z_1,\dots,z_{k-1}|w_1,\dots,w_{L-1}|I_1,\dots,I_{k-1}
|\lambda)
. \label{baserecursionwavefunction}
\end{align}

If $I_k \neq L$, the following factorizations hold for the base partition
functions (Figure \ref{picturebasepartitionfunctionfactorization}):
\begin{align}
&W_{L,k}(z_1,\dots,z_k|w_1,\dots,w_L|I_1,\dots,I_k|\lambda)
 \nonumber \\
=&\prod_{j=1}^k [z_j-w_L-\gamma]
W_{L-1,k}(z_1,\dots,z_k|w_1,\dots,w_{L-1}|I_1,\dots,I_k|\lambda).
\label{baserecursionwavefunction2}
\end{align}
\\
(4) The following evaluation holds for the case $k=1$, $I_1=L$:
\begin{align}
&
W_{L,1}(z_1|w_1,\dots,w_L|L|\lambda)
\nonumber \\
=&
\frac{[\gamma][z_1-w_L+\lambda_2-\lambda_1+\gamma(1-L)]}
{[\lambda_1-\lambda_2]}
\prod_{j=1}^{L-1} [z_1-w_j].
\label{baseinitialrecursion}
\end{align}
\end{proposition}
Proposition \ref{basewavefunctionizerginkorepin}
can be proved by the standard argument
using the graphical representations,
the dynamical Yang-Baxter relation and
the ice-rule for the six-vertex type models.

For example, 
Property (1) follows by inserting a completeness relation
between the space where the spectral variable $w_L$ is associated
and the space where the spectral variable $w_{L-1}$ is associated,
and split each base partition function into a sum
of products of factors.
The $w_L$-dependent factors for each summand 
can be computed by concentrating on the
dynamical $R$-matrices in the rightmost column
of the base partition functions, and has the following form
(Figure \ref{pictureforqpbase}):
\begin{align}
g_\ell(w_L)=
[z_\ell-w_L+\lambda_2-\lambda_1+(2k-L-\ell) \gamma]
\prod_{j=1}^{\ell-1} [z_j-w_L]
\prod_{j=\ell+1}^k [z_j-w_L-\gamma], \ \ \ \ell=1,\cdots,k.
\end{align}
One can easily calculate the quasi-periodicities
\begin{align}
&g_\ell(w_L+1)=(-1)^k g_\ell(w_L),
\\
&g_\ell(w_L+\tau)=(-1)^k \mathrm{exp}
\Bigg(-2 \pi i 
\Bigg(
k w_L-\sum_{i=1}^k z_i+\lambda_1-\lambda_2+\gamma (L-k)
\Bigg)
- \pi i k \tau \Bigg) g_\ell(w_L),
\end{align}
which is the same for all summands,
and one concludes the quasi-periodicities \eqref{qpbaseone}, \eqref{qpbasetwo}
hold.

Property (3) can be shown by using the graphical description of
the base partition functions.
When $I_k=L$, one finds that after the substitution $w_L=z_k-\gamma$,
the dynamical $R$-matrices at the lowest row and the rightmost column
get frozen, and the remaining unfrozen part is a smaller
base partition function
$W_{L-1,k-1}(z_1,\dots,z_{k-1}|w_1,\dots,w_{L-1}|I_1,\dots,I_{k-1}
|\lambda)$
(Figure \ref{picturebasepartitionfunctionrecursion}).
Multiplying $W_{L-1,k-1}(z_1,\dots,z_{k-1}|w_1,\dots,w_{L-1}|I_1,\dots,I_{k-1}
|\lambda)$ by the product of the weights of the dynamical $R$-matrices of the frozen part,
we get \eqref{baserecursionwavefunction}.

When $I_k \neq L$, one can easily see that the dynamical $R$-matrices
at the rightmost column are frozen, and the unfrozen part is
$W_{L-1,k}(z_1,\dots,z_k|w_1,\dots,w_{L-1}|I_1,\dots,I_k|\lambda)$
(Figure \ref{picturebasepartitionfunctionfactorization}).
Multiplying 
$W_{L-1,k}(z_1,\dots,z_k|w_1,\dots,w_{L-1}|I_1,\dots,I_k|\lambda)$ by
the product of the weights of the dynamical $R$-matrices
at the rightmost column, we have \eqref{baserecursionwavefunction2}. \\
\\

The next thing to do after proving Proposition \eqref{basewavefunctionizerginkorepin} is to show that the elliptic multivariable functions
$E_{L,k}(z_1,\dots,z_k|w_1,\dots,w_L|I_1,\dots,I_k|\lambda)$
\eqref{basesymmetricfunction}
satisfy all the properties in
Proposition \eqref{basewavefunctionizerginkorepin},
hence they are nothing but the explicit representations
for the base partition functions
$W_{L,k}(z_1,\dots,z_k|w_1,\dots,w_L|I_1,\dots,I_k|\lambda)$.

For example,
let us show Property (1) and (3)
for the case $I_k=L$.
We first note that each summand in
$E_{L,k}(z_1,\dots,z_k|w_1,\dots,w_L|I_1,\dots,I_k|\lambda)$
has a $w_L$-dependent factor
\begin{align}
f_{\sigma}(w_L)=[z_{\sigma(k)}-w_L+\lambda_2-\lambda_1+(2k-1-L) \gamma]
\prod_{i=1}^{k-1} [z_{\sigma(i)}-w_L-\gamma]. \label{summandbase}
\end{align}
The quasi-periodicites for these factors can be easily computed as
\begin{align}
&f_\sigma(w_L+1)=(-1)^k f_\sigma(w_L),
\\
&f_\sigma(w_L+\tau)=(-1)^k \mathrm{exp}
\Bigg(-2 \pi i 
\Bigg(
k w_L-\sum_{i=1}^k z_i+\lambda_1-\lambda_2+\gamma (L-k)
\Bigg)
- \pi i k \tau \Bigg) f_\sigma(w_L),
\end{align}
which is independent of $\sigma$, hence 
$E_{L,k}(z_1,\dots,z_k|w_1,\dots,w_L|I_1,\dots,I_k|\lambda)$
satisfy the required
quasi-periodicities \eqref{qpbaseone}, \eqref{qpbasetwo}
and are elliptic polynomials.

One also notes from \eqref{summandbase} that only the summands satisfying
$\sigma(k)=k$ survive after we set $w_L=z_k-\gamma$
in
$E_{L,k}(z_1,\dots,z_k|w_1,\dots,w_L|I_1,\dots,I_k|\lambda)$.
Then we find that
\eqref{basesymmetricfunction} can be rewritten as
\begin{align}
&E_{L,k}(z_1,\dots,z_k|w_1,\dots,w_L|I_1,\dots,I_k|\lambda)|_{w_L=z_k-\gamma}
\nonumber \\
=&\frac{[\gamma][\gamma]^{k-1}}
{[\lambda_1-\lambda_2+(1-k)\gamma]\prod_{j=1}^{k-1} [\lambda_1-\lambda_2+(1-j)\gamma]} \nonumber \\
\times&\sum_{\sigma \in S_{k-1}} 
\prod_{a=1}^{k-1}
\Bigg(
[z_{\sigma(a)}-w_{I_a}+\lambda_2-\lambda_1+(2a-1-I_a)\gamma]
\prod_{i=1}^{I_a-1}[z_{\sigma(a)}-w_i]
\prod_{i=I_a+1}^{L-1}[z_{\sigma(a)}-w_i-\gamma]
\Bigg) \nonumber \\
\times&[\lambda_2-\lambda_1+\gamma(2k-L)]
\prod_{j=1}^{L-1}[z_k-w_j] \prod_{i=1}^{k-1}[z_{\sigma(i)}-z_k]
\nonumber \\
\times&\prod_{a=1}^{k-1} \frac{[z_{\sigma(a)}-z_k+\gamma]}{[z_{\sigma(a)}-z_k]}
\prod_{1 \le a < b \le k-1}
\frac{[z_{\sigma(a)}-z_{\sigma(b)}+\gamma]}{[z_{\sigma(a)}-z_{\sigma(b)}]}
\nonumber \\
=&\frac{[\gamma][\lambda_2-\lambda_1+(2k-L)\gamma]}{[\lambda_1-\lambda_2+(1-k)\gamma]}
\prod_{j=1}^{k-1} [z_j-z_k+\gamma]
\prod_{j=1}^{L-1} [z_k-w_j]
\nonumber \\
&\times E_{L-1,k-1}(z_1,\dots,z_{k-1}|w_1,\dots,w_{L-1}|I_1,\dots,I_{k-1}
|\lambda).
\end{align}
This relation for the elliptic multivartiable functions
is exactly the same as the relation 
\eqref{baserecursionwavefunction}
for the base partition functions,
and hence property (3) for the case $I_k=L$ is shown.

Property (3) for the case $I_k \neq L$ can also be shown in
a simliar way.
The other properties are easy to check from the definition
of the elliptic functions \eqref{basesymmetricfunction}.

\end{proof}

\section{Partition functions of Foda-Manabe type associated with $E_{\tau,\eta}(gl_3)$}
We analyze the partition functions
$W(\{ {\bm z^{(1)}} \}, \{ {\bm z^{(2)}} \}
|\{ {\bm w^{(1)}} \}, \{ {\bm w^{(2)}} \}|\{k_1,k_2,L_1,L_2,
{\bm I} \}|\lambda)$ of Foda-Manabe type associated with $E_{\tau,\eta}(gl_3)$ in this section.

\subsection{Izergin-Korepin analysis}
In this subsection, we perform the Izergin-Korepin analysis
to determine the properties of the partition functions
$W(\{ {\bm z^{(1)}} \}, \{ {\bm z^{(2)}} \}
|\{ {\bm w^{(1)}} \}, \{ {\bm w^{(2)}} \}|\{k_1,k_2,L_1,L_2,{\bm I} \}|\lambda)$. We introduce the following notation
\begin{align}
c(k,j)=\# \{\ell | 1 \le \ell \le k, i_\ell^{(2)}=j \}.
\label{countingoperator}
\end{align}
The Korepin's Lemma corresponding to the partition functions
of Foda-Manabe type is given by the following proposition.

\begin{figure}[ht]
\includegraphics[width=12cm]{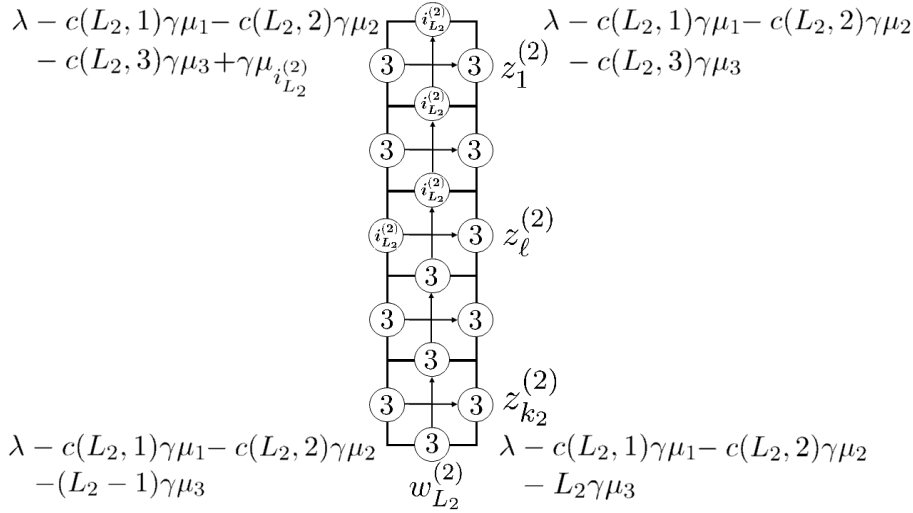}
\caption{
The states of the rightmost column of the upper region of the
partition functions
$W(\{ {\bm z^{(1)}} \}, \{ {\bm z^{(2)}} \}
|\{ {\bm w^{(1)}} \}, \{ {\bm w^{(2)}} \}
|\{k_1,k_2,L_1,L_2,{\bm I} \}|\lambda)$
giving non-zero contributions are graphically represented
as above.
The factors including $w_{L_2}^{(2)}$ all come from
the dynamical $R$-matrices in
this column and can be computed from the above graphical description.
We get
$\displaystyle h_\ell(w_{L_2}^{(2)})=
[z_\ell^{(2)}-w_{L_2}^{(2)}+\lambda_3-\lambda_{i_{L_2}^{(2)}}
+(c(L_2,i_{L_2}^{(2)})-c(L_2,3)-\ell) \gamma]
\prod_{j=1}^{\ell-1} [z_j^{(2)}-w_{L_2}^{(2)}]
\prod_{j=\ell+1}^{k_2} [z_j^{(2)}-w_{L_2}^{(2)}-\gamma], \ \ \ \ell=1,\cdots,k_2$.
}
\label{pictureforqpnested}
\end{figure}

\begin{figure}[ht]
\includegraphics[width=15cm]{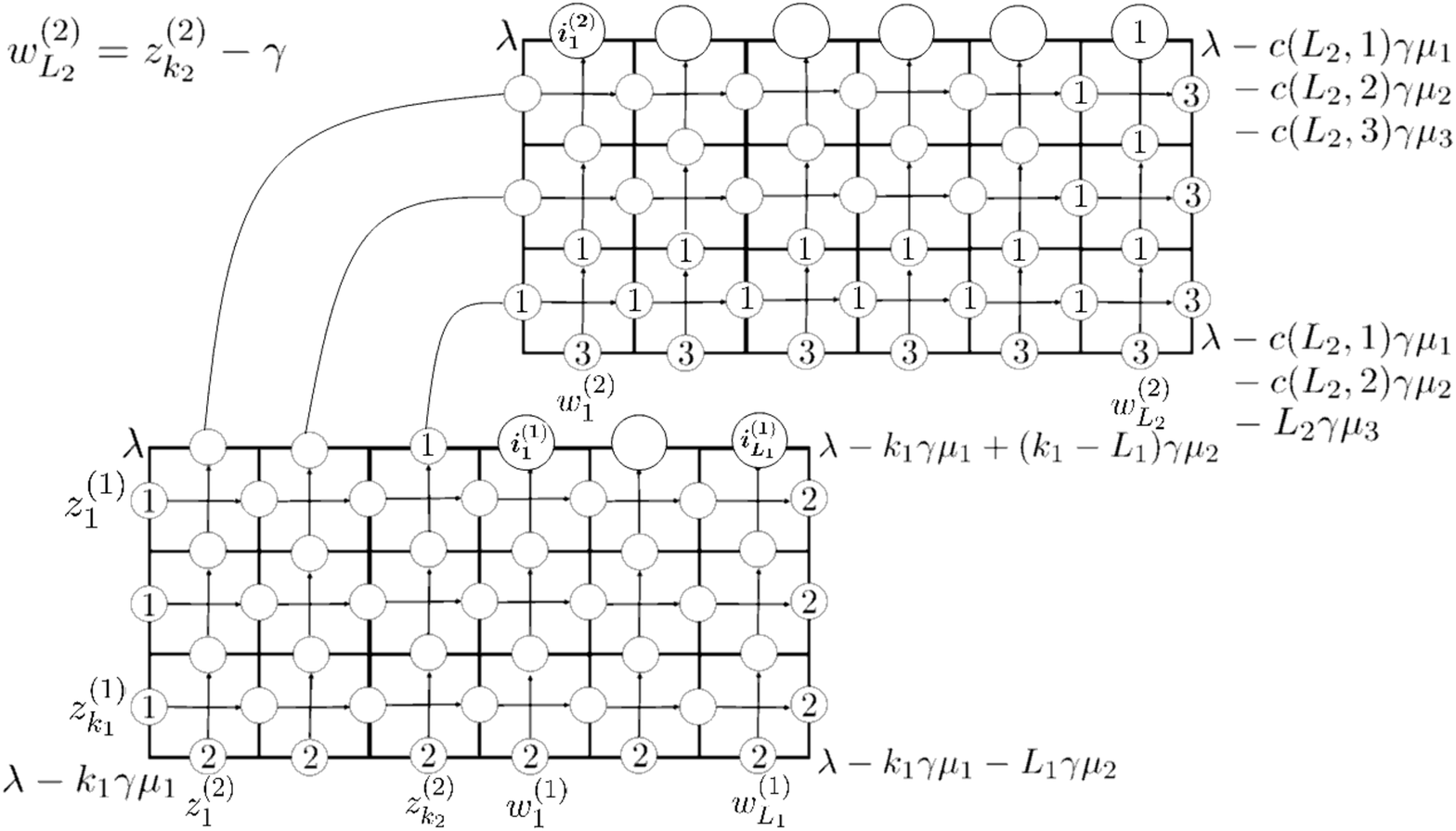}
\caption{The partition functions of Foda-Manabe type satisfying
$i_{L_2}^{(2)}=1$, evaluated at $w_{L_2}^{(2)}=z_{k_2}^{(2)}-\gamma$ \eqref{recursionwavefunction}. In the upper region, the dynamical $R$-matrices
at the rightmost column and the bottom row are all frozen.
}
\label{picturehigherrankpartitionfunctionrecursionone}
\end{figure}

\begin{figure}[ht]
\includegraphics[width=15cm]{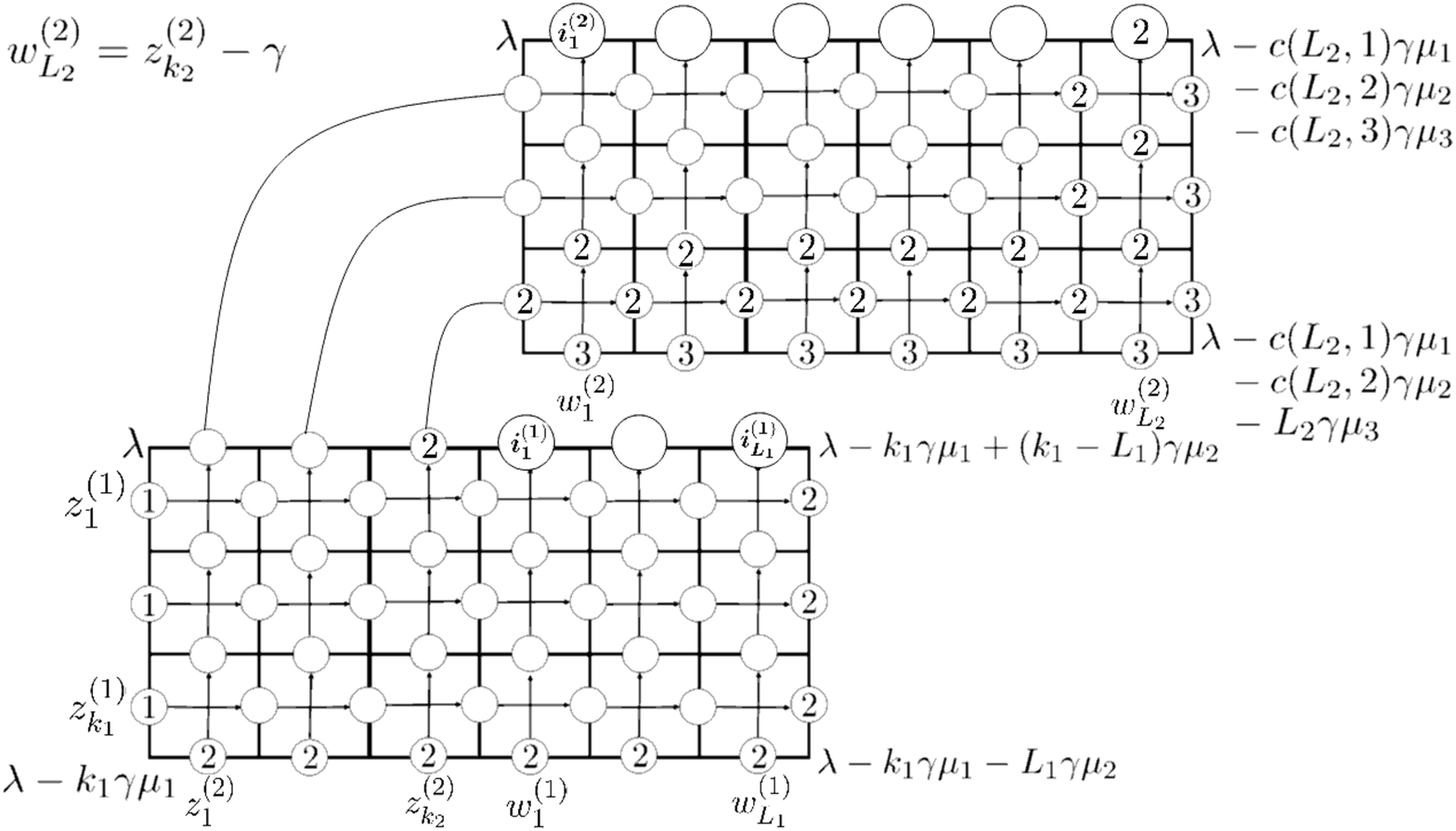}
\caption{The partition functions of Foda-Manabe type satisfying
$i_{L_2}^{(2)}=2$, evaluated at $w_{L_2}^{(2)}=z_{k_2}^{(2)}-\gamma$
\eqref{recursionwavefunction}. In the upper region, the dynamical $R$-matrices
at the rightmost column and the bottom row are all frozen.
}
\label{picturehigherrankpartitionfunctionrecursiontwo}
\end{figure}

\begin{figure}[ht]
\includegraphics[width=15cm]{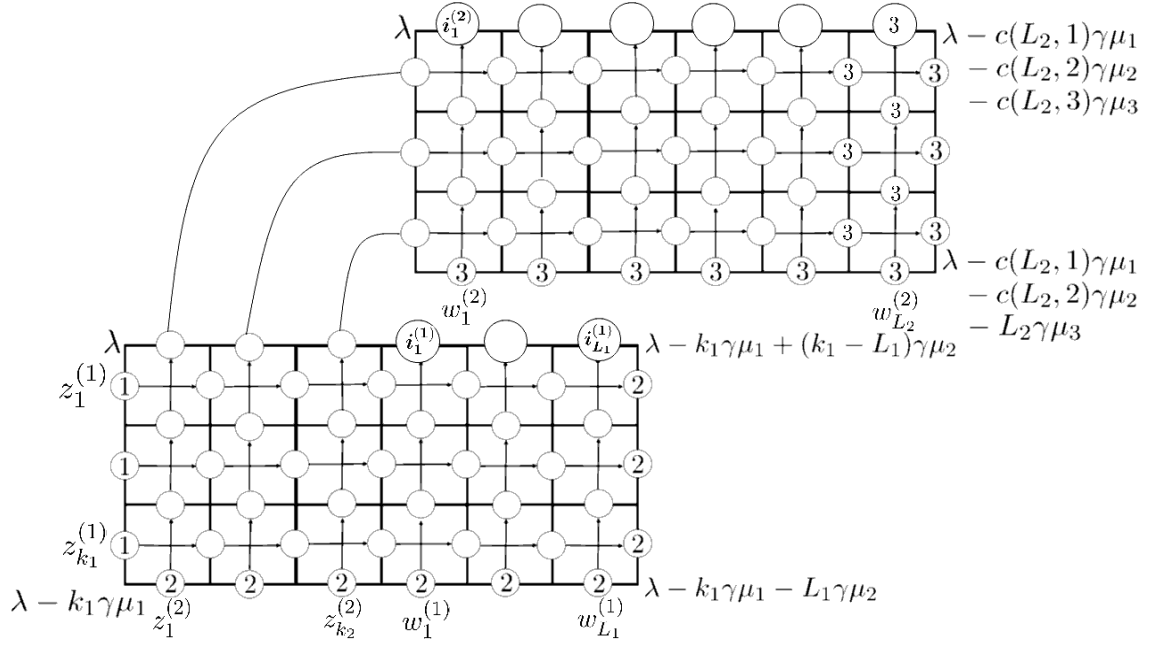}
\caption{The partition functions of Foda-Manabe type with $i_{L_2}^{(2)}=3$
\eqref{recursionwavefunction2}. The dynamical $R$-matrices at the
rightmost column in the upper region are all frozen.
}
\label{picturehigherrankpartitionfunctionfactorization}
\end{figure}

\begin{figure}[ht]
\includegraphics[width=15cm]{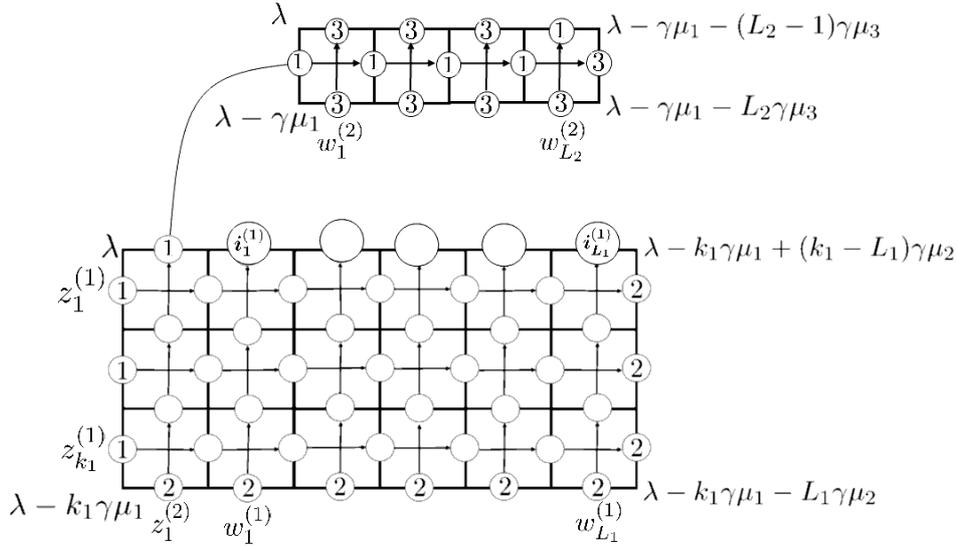}
\caption{The partition functions of Foda-Manabe type
with $k_2=1$ and $i^{(2)}_{L_2}=1$
\eqref{initialrecursion}.
One can easily see that the upper region is frozen,
and the lower region is a base partition function.
}
\label{picturehigherrankpartitionfunctioninitialone}
\end{figure}

\begin{figure}[ht]
\includegraphics[width=15cm]{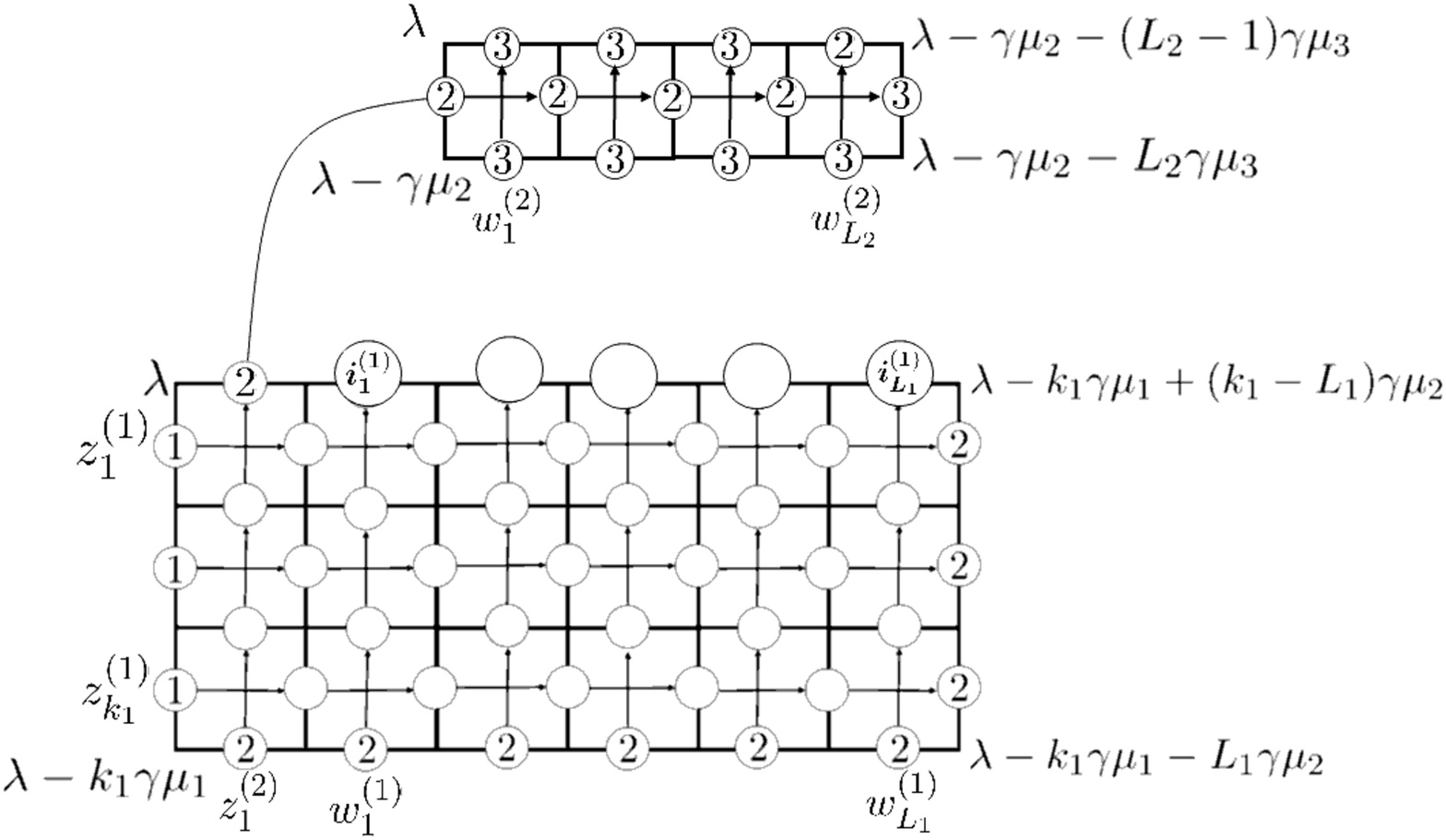}
\caption{The partition functions of Foda-Manabe type
with $k_2=1$ and $i^{(2)}_{L_2}=2$
\eqref{initialrecursion}.
One can easily see that the upper region is frozen,
and the lower region is a base partition function.
}
\label{picturehigherrankpartitionfunctioninitialtwo}
\end{figure}

\begin{proposition} 
\label{higherrankpartitionfunctionsizerginkorepin}
The partition functions of Foda-Manabe type
associated with $E_{\tau, \eta}(gl_3)$ \\
$W(\{ {\bm z^{(1)}} \}, \{ {\bm z^{(2)}} \}
|\{ {\bm w^{(1)}} \}, \{ {\bm w^{(2)}} \}
|\{k_1,k_2,L_1,L_2,{\bm I} \}|\lambda)$
possess the following properties. \\
\\
 (1) When $i^{(2)}_{L_2}$ satisfies $i^{(2)}_{L_2}=1$
or $i^{(2)}_{L_2}=2$, the partition
functions \\
$W(\{ {\bm z^{(1)}} \}, \{ {\bm z^{(2)}} \}
|\{ {\bm w^{(1)}} \}, \{ {\bm w^{(2)}} \}
|\{k_1,k_2,L_1,L_2,{\bm I} \}|\lambda)$
are elliptic polynomials of $w_{L_2}^{(2)}$ in $\Theta_{k_2}(\chi)$
with the following quasi-periodicities
\begin{align}
&W(\{ {\bm z^{(1)}} \}, \{ {\bm z^{(2)}} \}
|\{ {\bm w^{(1)}} \}, \{ {\bm w^{(2)}} \}|\{k_1,k_2,L_1,L_2,{\bm I} \}|\lambda)|_{w^{(2)}_{L_2} \longrightarrow w^{(2)}_{L_2}+1} \nonumber \\
=&(-1)^{k_2} W(\{ {\bm z^{(1)}} \}, \{ {\bm z^{(2)}} \}
|\{ {\bm w^{(1)}} \}, \{ {\bm w^{(2)}} \}|\{k_1,k_2,L_1,L_2,
{\bm I} \}|\lambda), \label{qpnestedone} \\
&W(\{ {\bm z^{(1)}} \}, \{ {\bm z^{(2)}} \}
|\{ {\bm w^{(1)}} \}, \{ {\bm w^{(2)}} \}|\{k_1,k_2,L_1,L_2,{\bm I} \}|\lambda)|_{w^{(2)}_{L_2} \longrightarrow w^{(2)}_{L_2}+\tau} \nonumber \\
=&
(-1)^{k_2} \mathrm{exp}
\Bigg(-2 \pi i 
\Bigg(
k_2 w_{L_2}^{(2)}-\sum_{i=1}^{k_2} z_i^{(2)}+\lambda_{i_{L_2}^{(2)}}-\lambda_3+\gamma (L_2-c(L_2,i^{(2)}_{L_2}))
\Bigg)
- \pi i k_2 \tau \Bigg)
\nonumber \\
&\times W(\{ {\bm z^{(1)}} \}, \{ {\bm z^{(2)}} \}
|\{ {\bm w^{(1)}} \}, \{ {\bm w^{(2)}} \}
|\{k_1,k_2,L_1,L_2,{\bm I} \}|\lambda) \label{qpnestedtwo}.
\end{align}
\\
 (2) The partition functions $W(\{ {\bm z^{(1)}} \}, \{ {\bm z^{(2)}} \}
|\{ {\bm w^{(1)}} \}, \{ {\bm w^{(2)}} \}
|\{k_1,k_2,L_1,L_2,{\bm I} \}|\lambda)$
are symmetric with respect to $z_{1}^{(2)},\dots,z_{k_2}^{(2)}$.
\\
\\
(3) If $i^{(2)}_{L_2}$ satisfies $i^{(2)}_{L_2}=1$
or $i^{(2)}_{L_2}=2$, the following recursive relations between the
partition functions hold (Figures \ref{picturehigherrankpartitionfunctionrecursionone}, \ref{picturehigherrankpartitionfunctionrecursiontwo}):
\begin{align}
&W(\{ {\bm z^{(1)}} \}, \{ {\bm z^{(2)}} \}
|\{ {\bm w^{(1)}} \}, \{ {\bm w^{(2)}} \}|\{k_1,k_2,L_1,L_2,{\bm I} \}|\lambda)
|_{w_{L_2}^{(2)}=z_{k_2}^{(2)}-\gamma}
\nonumber \\
=&\frac{[\gamma][\lambda_3-\lambda_{i^{(2)}_{L_2}}+\gamma(k_2-L_2+c(L_2,i^{(2)}_{L_2}))]}{[\lambda_{i_{L_2}^{(2)}}-\lambda_3+\gamma (1-c(L_2,i_{L_2}^{(2)}))]}
\prod_{j=1}^{k_2-1} [z_j^{(2)}-z_{k_2}^{(2)}+\gamma]
\prod_{j=1}^{L_2-1} [z_{k_2}^{(2)}-w_j^{(2)}]
\nonumber \\
\times&
W(\{ {\bm z^{(1)}} \}, \{ {z_1^{(2)}, \dots, z_{k_2-1}^{(2)} } \}
|\{ z_{k_2}^{(2)}, w_1^{(1)}, \dots, w_{L_1}^{(1)} \}, \{ w_1^{(2)}, \dots,
w_{L_2-1}^{(2)} \} \nonumber \\
&|\{k_1,k_2-1,L_1+1,L_2-1,{\bm J} \}|\lambda)
. \label{recursionwavefunction}
\end{align}
Here,
$W(\{ {\bm z^{(1)}} \}, \{ {z_1^{(2)}, \dots, z_{k_2-1}^{(2)} } \}
|\{ z_{k_2}^{(2)}, w_1^{(1)}, \dots, w_{L_1}^{(1)} \}, \{ w_1^{(2)}, \dots,
w_{L_2-1}^{(2)} \}
|\{k_1,k_2-1,L_1+1,L_2-1,{\bm J} \}|\lambda)$
are partition functions of size $\{ k_1,k_2-1,L_1+1,L_2-1 \}$
whose configuration of colors are labeled by a
set 
${\bm J}:=\{{\bm J_{k_1}^{(1)}}, {\bm J_{k_2-1}^{(2)}},
 \widehat{{\bm J}}_{(k_2-1)+(L_1+1)}^{(2)}, \widehat{{\bm J}}_{L_2-1}^{(3)} \}$
where the subsets
${\bm J_{k_1}^{(1)}}$, ${\bm J_{k_2-1}^{(2)}}$,
$\widehat{{\bm J}}_{(k_2-1)+(L_1+1)}^{(2)}$, $\widehat{{\bm J}}_{L_2-1}^{(3)}$ are given by
${\bm J_{k_1}^{(1)}}={\bm I_{k_1}^{(1)}}$, ${\bm J_{k_2-1}^{(2)}}
={\bm I_{k_2}^{(2)}} \backslash \{ L_2 \}$,
$\widehat{{\bm J}}_{(k_2-1)+(L_1+1)}^{(2)}=
\widehat{{\bm I}}_{k_2+L_1}^{(2)}
$, $\widehat{{\bm J}}_{L_2-1}^{(3)}=\{1,\dots,L_2-1 \}$
satisfying the following inclusion relations
\begin{align}
&{\bm J}_{k_1}^{(1)} \subset \widehat{{\bm J}}_{(k_2-1)+(L_1+1)}^{(2)}
:={\bm J_{k_2-1}^{(2)}} \cup \{ L_2,\dots,L_2+L_1 \}, \\
&{\bm J_{k_2-1}^{(2)}} \subset \widehat{{\bm J}}_{L_2-1}^{(3)}:=\{1,\dots,L_2-1 \}.
\end{align}

If $i^{(2)}_{L_2}=3$, the following factorizations hold for the 
partition functions (Figure \ref{picturehigherrankpartitionfunctionfactorization}):
\begin{align}
&W(\{ {\bm z^{(1)}} \}, \{ {\bm z^{(2)}} \}
|\{ {\bm w^{(1)}} \}, \{ {\bm w^{(2)}} \}|\{k_1,k_2,L_1,L_2,{\bm I} \}|\lambda)
 \nonumber \\
=&
\prod_{j=1}^{k_2} [z_j^{(2)}-w_{L_2}^{(2)}-\gamma] \nonumber \\
\times&
W(\{ {\bm z^{(1)}} \}, \{ {\bm z^{(2)}} \}
|\{ {\bm w^{(1)}} \}, \{ w_1^{(2)},\dots, w_{L_2-1}^{(2)} \}|\{k_1,k_2,L_1,L_2-1,{\bm K} \}|\lambda).
\label{recursionwavefunction2}
\end{align}
Here,
$W(\{ {\bm z^{(1)}} \}, \{ {\bm z^{(2)}} \}
|\{ {\bm w^{(1)}} \}, \{ w_1^{(2)},\dots, w_{L_2-1}^{(2)} \}|\{k_1,k_2,L_1,L_2-1,{\bm K} \}|\lambda)$
are partition functions of size $\{ k_1,k_2,L_1,L_2-1 \}$
whose configuration of colors are labeled by a
set 
${\bm K}:=\{{\bm K_{k_1}^{(1)}}, {\bm K_{k_2}^{(2)}},
 \widehat{{\bm K}}_{k_2+L_1}^{(2)}, \widehat{{\bm K}}_{L_2-1}^{(3)} \}$
where the subsets
${\bm K_{k_1}^{(1)}}$, ${\bm K_{k_2}^{(2)}}$,
$\widehat{{\bm K}}_{k_2+L_1}^{(2)}$, $\widehat{{\bm K}}_{L_2-1}^{(3)}$
are given by
${\bm K_{k_1}^{(1)}}=
{\bm I_{k_1}^{(1)}}|_{L_2+1 \rightarrow L_2,\dots,L_2+L_1 \rightarrow L_2+L_1-1}$, ${\bm K_{k_2}^{(2)}}={\bm I_{k_2}^{(2)}}$,
$\widehat{{\bm K}}_{k_2+L_1}^{(2)}={\bm I_{k_2}^{(2)}} \cup \{L_2,\dots,L_2+L_1-1 \}$, $\widehat{{\bm K}}_{L_2-1}^{(3)}=\{1, \dots, L_2-1 \}$
satisfying the following inclusion relations
\begin{align}
&{\bm K}_{k_1}^{(1)} \subset \widehat{{\bm K}}_{k_2+L_1}^{(2)}
:={\bm K_{k_2}^{(2)}} \cup \{ L_2,\dots,L_2+L_1-1 \}, \\
&{\bm K_{k_2}^{(2)}} \subset \widehat{{\bm K}}_{L_2-1}^{(3)}:=\{1,\dots,L_2-1 \}.
\end{align}
\\
(4) When $k_2=1$ and $i^{(2)}_{L_2}$ satisfies $i^{(2)}_{L_2}=1$
or $i^{(2)}_{L_2}=2$,
the following evaluation holds
(Figures \ref{picturehigherrankpartitionfunctioninitialone},
\ref{picturehigherrankpartitionfunctioninitialtwo})
:
\begin{align}
&
W(\{ {\bm z^{(1)}} \}, \{ {\bm z^{(2)}} \}
|\{ {\bm w^{(1)}} \}, \{ {\bm w^{(2)}} \}|\{k_1,1,L_1,L_2,{\bm I} \}|\lambda)
\nonumber \\
=&
\frac{[\gamma][z_1^{(2)}-w_{L_2}^{(2)}+\lambda_3-\lambda_{i_{L_2}^{(2)}}-\gamma(L_2-1)]}{[\lambda_{i_{L_2}^{(2)}}-\lambda_3]}
\prod_{j=1}^{L_2-1} [z_1^{(2)}-w_j^{(2)}] \nonumber \\
\times&
W_{L_1+1,k_1}(z_1^{(1)},\dots,z_{k_1}^{(1)}|
z_1^{(2)},w_1^{(1)},\dots,w_{L_1}^{(1)}|I_1^{(1)}+1-L_2,\dots,I_{k_1}^{(1)}+1-L_2|\lambda).
\label{initialrecursion}
\end{align}
\end{proposition}

Let us give some remarks on
Proposition \ref{higherrankpartitionfunctionsizerginkorepin},
which corresponds to the Korepin's lemma for the case of
higher rank partition functions.
An idea to anlayze partition functions
which goes back to Korepin \cite{Ko} is to construct relations
between partition functions of different sizes.
In the present case, the partition functions 
of type $\{k_1,k_2,L_1,L_2,
{\bm I} \}$ is connected with
other smaller partition functions
(which have smaller $k_1+k_2+L_1+L_2$),
and which smaller partition functions are connected
depend on the color $i_{L_2}^{(2)}$ in the northeast corner.
When $i_{L_2}^{(2)}=1$ or $i^{(2)}_{L_2}=2$,
partition functions of type
$\{k_1,k_2,L_1,L_2,
{\bm I} \}$
are connected with the ones of type
$\{k_1,k_2-1,L_1+1,L_2-1,
{\bm J} \}$,
When $i_{L_2}^{(2)}=3$,
the partition functions of type
$\{k_1,k_2,L_1,L_2,
{\bm I} \}$
are reduced to the ones of type $\{k_1,k_2,L_1,L_2-1,{\bm K} \}$.
The case $k_2=1$ and $i^{(2)}_{L_2}=1$ or $i^{(2)}_{L_2}=2$
correspond to the initial partition functions for this recursion,
and are essentially given by the base partition functions
$W_{L_1+1,k_1}(z_1^{(1)},\dots,z_{k_1}^{(1)}|
z_1^{(2)},w_1^{(1)},\dots,w_{L_1}^{(1)}|I_1^{(1)}+1-L_2,\dots,I_{k_1}^{(1)}+1-L_2|\lambda)$ which are analyzed in the previous section.

\begin{proof}
The proof of the Izergin-Korepin analysis is
basically the same with the case for the
base partition functions.

Property (1) can be shown by
inserting a completeness relation
between the space where the spectral variable $w_{L_2}^{(2)}$ is associated
and the space where the spectral variable $w_{L_2-1}^{(2)}$ is associated
to split the partition functions into a sum of factors.
The $w_{L_2}^{(2)}$-dependent factors for each summand has the following form
(Figure \ref{pictureforqpnested}):
\begin{align}
h_\ell(w_{L_2}^{(2)})=&
[z_\ell^{(2)}-w_{L_2}^{(2)}+\lambda_3-\lambda_{i_{L_2}^{(2)}}
+(c(L_2,i_{L_2}^{(2)})-c(L_2,3)-\ell) \gamma] \nonumber \\
\times&\prod_{j=1}^{\ell-1} [z_j^{(2)}-w_{L_2}^{(2)}]
\prod_{j=\ell+1}^{k_2} [z_j^{(2)}-w_{L_2}^{(2)}-\gamma], \ \ \ \ell=1,\cdots,k_2.
\end{align}
Calculating the quasi-periodicities of $h_\ell(w_{L_2}^{(2)})$, we get
\begin{align}
h_\ell(w_{L_2}^{(2)}+1)&=(-1)^{k_2} h_\ell(w_{L_2}^{(2)}),
\\
h_\ell(w_{L_2}^{(2)}+\tau)&=(-1)^{k_2} \mathrm{exp}
\Bigg(-2 \pi i 
\Bigg(
k_2 w_{L_2}^{(2)}-\sum_{i=1}^{k_2} z_i^{(2)}
+\lambda_{i_{L_2}^{(2)}}
-\lambda_3
\nonumber \\
&+\gamma (k_2+c(L_2,3)-c(L_2,i_{L_2}^{(2)}))
\Bigg)
- \pi i k_2 \tau \Bigg) h_\ell(w_{L_2}^{(2)}),
\end{align}
and one finds that they are all the same for all summands,
hence we get \eqref{qpnestedone} and \eqref{qpnestedtwo}
(also note that $c(L_2,3)=L_2-k_2$).

Property (2) follows from the standard railroad
argument using the dynamical Yang-Baxter relation.
Note that the height variables at the northwest corners
in the upper region and the lower region are both fixed to $\lambda$
so that the railroad argument can be applied.

Property (3) can be shown by using the graphical representation
for the partition functions.
We look at the upper region where the spectral parameters
$w_1^{(2)},\dots,w_{L_2}^{(2)}$ are associated.
When $i^{(2)}_{L_2}=1$
or $i^{(2)}_{L_2}=2$, one can see that
after the substitution $w_{L_2}^{(2)}=z_{k_2}^{(2)}-\gamma$,
the dynamical $R$-matrices at the bottom row and the rightmost column
in the upper region are frozen
(Figures \ref{picturehigherrankpartitionfunctionrecursionone},
\ref{picturehigherrankpartitionfunctionrecursiontwo}).
The  product of the matrix elements of the dynamical $R$-matrices of the frozen part gives the factor
\begin{align}
\frac{[\gamma][\lambda_3-\lambda_{i^{(2)}_{L_2}}+\gamma(k_2-L_2+c(L_2,i^{(2)}_{L_2}))]}{[\lambda_{i_{L_2}^{(2)}}-\lambda_3+\gamma (1-c(L_2,i_{L_2}^{(2)}))]}
\prod_{j=1}^{k_2-1} [z_j^{(2)}-z_{k_2}^{(2)}+\gamma]
\prod_{j=1}^{L_2-1} [z_{k_2}^{(2)}-w_j^{(2)}],
\label{factorhigherpartitionfunctionone}
\end{align}
in the right hand side of \eqref{recursionwavefunction}.
Next we look at the unfrozen part and we view this as
a partition function of a smaller size $\{k_1,k_2-1,L_1+1,L_2-1 \}$.
The configuration of colors are encoded into
a set which we denote by
${\bm J}:=\{{\bm J_{k_1}^{(1)}}, {\bm J_{k_2-1}^{(2)}},
 \widehat{{\bm J}}_{(k_2-1)+(L_1+1)}^{(2)},
\widehat{{\bm J}}_{L_2-1}^{(3)} \}$.
The set ${\bm J }$ is related with the set
${\bm I}:=\{{\bm I_{k_1}^{(1)}}, {\bm I_{k_2}^{(2)}},
 \widehat{{\bm I}}_{k_2+L_1}^{(2)}, \widehat{{\bm I}}_{L_2}^{(3)} \}$
for the original partition functions,
and we can see from the encoding rule of
the configuration of colors into sets explained in section 3
that the subsets of ${\bm J}$ and ${\bm I}$
are related by the following relations:
${\bm J_{k_1}^{(1)}}={\bm I_{k_1}^{(1)}}$, ${\bm J_{k_2-1}^{(2)}}
={\bm I_{k_2}^{(2)}} \backslash \{ L_2 \}$,
$\widehat{{\bm J}}_{(k_2-1)+(L_1+1)}^{(2)}=
\widehat{{\bm I}}_{k_2+L_1}^{(2)}
$, $\widehat{{\bm J}}_{L_2-1}^{(3)}=\{1,\dots,L_2-1 \}$.
Hence, we conclude that the original partition functions
$W(\{ {\bm z^{(1)}} \}, \{ {\bm z^{(2)}} \}
|\{ {\bm w^{(1)}} \}, \{ {\bm w^{(2)}} \}
|\{k_1,k_2,L_1,L_2,{\bm I} \}|\lambda)$
evaluated at $w_{L_2}^{(2)}=z_{k_2}^{(2)}-\gamma$
is given by the product of the factor
\eqref{factorhigherpartitionfunctionone}
and the smaller partition functions
$
W(\{ {\bm z^{(1)}} \}, \{ {z_1^{(2)}, \dots, z_{k_2-1}^{(2)} } \}
|\{ z_{k_2}^{(2)}, w_1^{(1)}, \dots, w_{L_1}^{(1)} \}, \{ w_1^{(2)}, \dots,
w_{L_2-1}^{(2)} \}
|\{k_1,k_2-1,L_1+1,L_2-1,{\bm J} \}|\lambda)
$, i.e. we get \eqref{recursionwavefunction}.

We can also show \eqref{recursionwavefunction2} with the help
of the graphical description.
When $i^{(2)}_{L_2}=3$,
one can see that the dynamical $R$-matrices at the rightmost column
in the upper region are frozen (Figure \ref{picturehigherrankpartitionfunctionfactorization}), and the matrix elements of the dynamical $R$-matrices
of this column gives the factor
$\prod_{j=1}^{k_2} [z_j-w_{L_2}^{(2)}-\gamma]$.
Peeling off the column, we get a smaller partition function 
of size $\{ k_1,k_2,L_1,L_2-1 \}$,
denoted by
$W(\{ {\bm z^{(1)}} \}, \{ {\bm z^{(2)}} \}
|\{ {\bm w^{(1)}} \}, \{ w_1^{(2)},\dots, w_{L_2-1}^{(2)} \}|\{k_1,k_2,L_1,L_2-1,{\bm K} \}|\lambda)$.
One can see that the set 
${\bm K}:=\{{\bm K_{k_1}^{(1)}}, {\bm K_{k_2}^{(2)}},
 \widehat{{\bm K}}_{k_2+L_1}^{(2)}, \widehat{{\bm K}}_{L_2-1}^{(3)} \}$,
which encodes the configuration of colors
for the smaller partition function,
is related with the set
${\bm I}:=\{{\bm I_{k_1}^{(1)}}, {\bm I_{k_2}^{(2)}},
 \widehat{{\bm I}}_{k_2+L_1}^{(2)}, \widehat{{\bm I}}_{L_2}^{(3)} \}$
for the original partition functions by the following relations:
${\bm K_{k_1}^{(1)}}=
{\bm I_{k_1}^{(1)}}|_{L_2+1 \rightarrow L_2,\dots,L_2+L_1 \rightarrow L_2+L_1-1}$, ${\bm K_{k_2}^{(2)}}={\bm I_{k_2}^{(2)}}$,
$\widehat{{\bm K}}_{k_2+L_1}^{(2)}={\bm I_{k_2}^{(2)}} \cup \{L_2,\dots,L_2+L_1-1 \}$, $\widehat{{\bm K}}_{L_2-1}^{(3)}=\{1, \dots, L_2-1 \}$.
Hence we find that
the original partition functions are given by multiplying
the factor $\prod_{j=1}^{k_2} [z_j-w_{L_2}^{(2)}-\gamma]$ by
$W(\{ {\bm z^{(1)}} \}, \{ {\bm z^{(2)}} \}
|\{ {\bm w^{(1)}} \}, \{ w_1^{(2)},\dots, w_{L_2-1}^{(2)} \}|\{k_1,k_2,L_1,L_2-1,{\bm K} \}|\lambda)$, i.e. we get \eqref{recursionwavefunction2}.

Property (4), which corresponds to the initial conditions of the recursion,
can also be shown by graphical descriptions.
When $k_2=1$ and $i^{(2)}_{L_2}=1$ or $i^{(2)}_{L_2}=2$,
one can see that the dynamical $R$-matrices in the upper region are all frozen
(Figures \ref{picturehigherrankpartitionfunctioninitialone},
\ref{picturehigherrankpartitionfunctioninitialtwo}), and the product of
the matrix elements of the dynamical $R$-matrices gives the factor
\begin{align}
\frac{[\gamma][z_1^{(2)}-w_{L_2}^{(2)}+\lambda_3-\lambda_{i_{L_2}^{(2)}}-\gamma(L_2-1)]}{[\lambda_{i_{L_2}^{(2)}}-\lambda_3]}
\prod_{j=1}^{L_2-1} [z_1^{(2)}-w_j^{(2)}]. \label{factorforinitial}
\end{align}
The unfrozen part is the lower region,
which is nothing but the base partition function.
One can see that using the elements of the set ${\bm I_{k_1}^{(1)}}
=\{ I_1^{(1)}, \dots, I_{k_1}^{(1)} \}$ which label the configuration
of colors for the original partition functions of Foda-Manabe type,
the base partition function which appears as the unfrozen part is
$
W_{L_1+1,k_1}(z_1^{(1)},\dots,z_{k_1}^{(1)}|
z_1^{(2)},w_1^{(1)},\dots,w_{L_1}^{(1)}|I_1^{(1)}+1-L_2,\dots,I_{k_1}^{(1)}+1-L_2|\lambda).
$ Hence, we find the original partition functions are given by the product of
\eqref{factorforinitial} and
$
W_{L_1+1,k_1}(z_1^{(1)},\dots,z_{k_1}^{(1)}|
z_1^{(2)},w_1^{(1)},\dots,w_{L_1}^{(1)}|I_1^{(1)}+1-L_2,\dots,I_{k_1}^{(1)}+1-L_2|\lambda)
$ \eqref{initialrecursion}.

\end{proof}

\subsection{Elliptic multivariable functions}
In this subsection,
we show the elliptic multivariable functions
defined below give the explicit representations
for the partition functions of Foda-Manabe type
by showing that they satisfy all the
properties in Proposition \ref{higherrankpartitionfunctionsizerginkorepin}
which $W(\{ {\bm z^{(1)}} \}, \{ {\bm z^{(2)}} \}
|\{ {\bm w^{(1)}} \}, \{ {\bm w^{(2)}} \}|\{k_1,k_2,L_1,L_2,
{\bm I} \}|\lambda)$ possess.

\begin{definition}
We define the following elliptic multivariable function \\
$E(\{ {\bm z^{(1)}} \}, \{ {\bm z^{(2)}} \}
|\{ {\bm w^{(1)}} \}, \{ {\bm w^{(2)}} \}|\{k_1,k_2,L_1,L_2,
{\bm I} \}|\lambda)$ which depend on two sets of symmetric variables
$\{ {\bm z^{(1)}} \}$, $\{ {\bm z^{(2)}} \}$,
two sets of complex parameters
$\{ {\bm w^{(1)}} \}$, $\{ {\bm w^{(2)}} \}$,
complex parameters $\gamma$, $\lambda_1$, $\lambda_2$, $\lambda_3$
and a set $\{k_1,k_2,L_1,L_2,{\bm I} \}$,
${\bm I}=\{{\bm I_{k_1}^{(1)}}, {\bm I_{k_2}^{(2)}},
\widehat{{\bm I}}_{k_2+L_1}^{(2)}, \widehat{{\bm I}}_{L_2}^{(3)} \}$
which is equivalent to ``the configuration of colors''
$i_1^{(1)},\dots,\dots,i_{L_1}^{(1)} \in \{1,2 \}$,
$i_1^{(2)},\dots,\dots,i_{L_2}^{(2)} \in \{1,2,3 \}$,
\begin{align}
&E(\{ {\bm z^{(1)}} \}, \{ {\bm z^{(2)}} \}
|\{ {\bm w^{(1)}} \}, \{ {\bm w^{(2)}} \}|\{k_1,k_2,L_1,L_2,{\bm I} \}|\lambda)
\nonumber \\
=&[\gamma]^{k_1+k_2}
\sum_{\sigma_1 \in S_{k_1}} \sum_{\sigma_2 \in S_{k_2}}
\prod_{a=1}^{k_1}
\Bigg(
\prod_{i=1}^{\widetilde{I_a^{(1)}}-1}
[z_{\sigma_1(a)}^{(1)}-m_i^{k_2,L_1}
(\{ {\bm z_{\sigma_2}^{(2)}} \}| \{ {\bm w^{(1)}} \})
] \nonumber \\
\times&
\frac{[z_{\sigma_1(a)}^{(1)}-
m_{\widetilde{I_a^{(1)}}}^{k_2,L_1} (\{ {\bm z_{\sigma_2}^{(2)}} \}| \{ {\bm w^{(1)}} \})
+\lambda_2-\lambda_1+\gamma(2a-1-\widetilde{I_a^{(1)}})
]}
{[\lambda_1-\lambda_2+(1-a)\gamma]} \nonumber \\
\times&\prod_{i={\widetilde{I_a^{(1)}}+1}}^{k_2+L_1}
[z_{\sigma_1(a)}^{(1)}-m_i^{k_2,L_1}
(\{ {\bm z_{\sigma_2}^{(2)}} \}| \{ {\bm w^{(1)}} \})-\gamma
]
\Bigg) \nonumber \\
\times&\prod_{1 \le a < b \le k_1} 
\frac{[z_{\sigma_1(a)}^{(1)}-z_{\sigma_1(b)}^{(1)}+\gamma]}
{[z_{\sigma_1(a)}^{(1)}-z_{\sigma_1(b)}^{(1)}]}
\prod_{a=1}^{k_2} \Bigg(
\prod_{i=1}^{I_a^{(2)}-1} [z_{\sigma_2(a)}^{(2)}-w_i^{(2)}]
\nonumber \\
\times&\frac{
[z_{\sigma_2(a)}^{(2)}-w_{I_a^{(2)}}^{(2)}+\lambda_3
-\lambda_{i_{I_a^{(2)}}^{(2)}}+(a-I_a^{(2)}-1+c(I_a^{(2)},i_{I_a^{(2)}}^{(2)}))\gamma]
}
{
[\lambda_{i_{I_a^{(2)}}^{(2)}}-\lambda_3
+(1-c(I_a^{(2)},i_{I_a^{(2)}}^{(2)}))\gamma]
}
\nonumber \\
\times&\prod_{i=I_a^{(2)}+1}^{L_2} [z_{\sigma_2(a)}^{(2)}-w_i^{(2)}-\gamma]
\Bigg)
\prod_{1 \le a < b \le k_2} 
\frac{[z_{\sigma_2(a)}^{(2)}-z_{\sigma_2(b)}^{(2)}+\gamma]}
{[z_{\sigma_2(a)}^{(2)}-z_{\sigma_2(b)}^{(2)}]},
\label{multivariablefunction}
\end{align}
where $m_i^{k_2,L_1}(\{ {\bm z^{(2)}} \}|\{ {\bm w^{(1)}} \})$
and $c(k,j)$ are given by
\begin{align}
m_i^{k_2,L_1}(\{ {\bm z^{(2)}} \}|\{ {\bm w^{(1)}} \})
= \begin{cases}
    z_i^{(2)} & 1 \le i \le k_2 \\
    w_{i-k_2}^{(1)} & k_2+1 \le i \le k_2+L_1
  \end{cases}, \label{definitionofm}
\end{align}
and
\begin{align}
c(k,j)=\# \{\ell | 1 \le \ell \le k, i_\ell^{(2)}=j \}.
\end{align}
The relation between the Foda-Manabe label $\{k_1,k_2,L_1,L_2,{\bm I} \}$,
${\bm I}=\{{\bm I_{k_1}^{(1)}}, {\bm I_{k_2}^{(2)}},
\widehat{{\bm I}}_{k_2+L_1}^{(2)}, \widehat{{\bm I}}_{L_2}^{(3)} \}$
and ``the configuration of colors''
$i_1^{(1)},\dots,\dots,i_{L_1}^{(1)} \in \{1,2 \}$,
$i_1^{(2)},\dots,\dots,i_{L_2}^{(2)} \in \{1,2,3 \}$,
and the induced label $\widetilde{{\bm I_{k_1}^{(1)}}}$
which is induced from the map
$\displaystyle
{\bm I}_{k_1}^{(1)} \subset \widehat{{\bm I}}_{k_2+L_1}^{(2)}
\longrightarrow \widetilde{{\bm I_{k_1}^{(1)}}} \subset
\{1,\dots,k_2+L_1 \}
$, are explained in section 3.
\end{definition}

\begin{theorem}
The partition functions
$W(\{ {\bm z^{(1)}} \}, \{ {\bm z^{(2)}} \}
|\{ {\bm w^{(1)}} \}, \{ {\bm w^{(2)}} \}
|\{k_1,k_2,L_1,L_2,{\bm I} \}|\lambda)$ of Foda-Manabe type
associated with $E_{\tau,\gamma}(gl_3)$
are explicitly expressed as the multivariable elliptic symmetric functions
$E(\{ {\bm z^{(1)}} \}, \{ {\bm z^{(2)}} \}
|\{ {\bm w^{(1)}} \}, \{ {\bm w^{(2)}} \}|\{k_1,k_2,L_1,L_2,{\bm I} \}|\lambda)$,
\begin{align}
&W(\{ {\bm z^{(1)}} \}, \{ {\bm z^{(2)}} \}
|\{ {\bm w^{(1)}} \}, \{ {\bm w^{(2)}} \}|\{k_1,k_2,L_1,L_2,{\bm I} \}|\lambda)
\nonumber \\
=&E(\{ {\bm z^{(1)}} \}, \{ {\bm z^{(2)}} \}
|\{ {\bm w^{(1)}} \}, \{ {\bm w^{(2)}} \}|\{k_1,k_2,L_1,L_2,{\bm I} \}|\lambda).
\end{align}

\end{theorem}

\begin{proof}
We show that the functions $E(\{ {\bm z^{(1)}} \}, \{ {\bm z^{(2)}} \}
|\{ {\bm w^{(1)}} \}, \{ {\bm w^{(2)}} \}
|\{k_1,k_2,L_1,L_2,{\bm I} \}|\lambda)$ satisfy all the
properties in
Proposition \ref{higherrankpartitionfunctionsizerginkorepin}
which the partition functions of Foda-Manabe type
$W(\{ {\bm z^{(1)}} \}, \{ {\bm z^{(2)}} \}
|\{ {\bm w^{(1)}} \}, \{ {\bm w^{(2)}} \}
|\{k_1,k_2,L_1,L_2,{\bm I} \}|\lambda)$ satisfy.

Properties (2) and (4) are easy to check from the definition
of the elliptic multivariable functions
$E(\{ {\bm z^{(1)}} \}, \{ {\bm z^{(2)}} \}
|\{ {\bm w^{(1)}} \}, \{ {\bm w^{(2)}} \}
|\{k_1,k_2,L_1,L_2,{\bm I} \}|\lambda)$
\eqref{multivariablefunction}.

Let us show Property (1) and
\eqref{recursionwavefunction} in Property (3).
When $i_{L_2}^{(2)}=1$ or $i_{L_2}^{(2)}=2$,
we note $I_{k_2}^{(2)}=L_2$
since $L_2 \in {\bm I_{k_2}^{(2)}}$.
From this fact, one finds that each summand
in \eqref{multivariablefunction} contains the following product of factors
\begin{align}
&f_{\sigma_2}(w_{L_2}^{(2)}) \nonumber \\
&=[z_{\sigma_2(k_2)}^{(2)}-w_{L_2}^{(2)}+\lambda_3-\lambda_{i_{L_2}^{(2)}}+\gamma(k_2-L_2-1+c(L_2,i_{L_2}^{(2)}))]
\prod_{a=1}^{k_2-1}
[z_{\sigma_2(a)}^{(2)}-w_{L_2}^{(2)}-\gamma], \label{factorinsummand}
\end{align}
from which all the $w_{L_2}^{(2)}$-dependence comes.
One can easily compute the quasi-periodicities
for $f_{\sigma_2}(w_{L_2}^{(2)})$
\begin{align}
&f_{\sigma_2}(w_{L_2}^{(2)}+1)=(-1)^{k_2}
f_{\sigma_2}(w_{L_2}^{(2)}),
\nonumber \\
&f_{\sigma_2}(w_{L_2}^{(2)}+\tau) \nonumber \\
=&(-1)^{k_2} \mathrm{exp}
\Bigg(-2 \pi i 
\Bigg(
k_2 w_{L_2}^{(2)}-\sum_{i=1}^{k_2} z_i^{(2)}+\lambda_{i_{L_2}^{(2)}}-\lambda_3+\gamma (L_2-c(L_2,i^{(2)}_{L_2}))
\Bigg)
- \pi i k_2 \tau \Bigg) f_{\sigma_2}(w_{L_2}^{(2)}). \nonumber
\end{align}
We find the quasi-periodicities are independent of $\sigma_2$,
and from this explicit expression,
one concludes that the elliptic multivariable functions
satisfy the same quasi-periodicities with
the partition functions \eqref{qpnestedone} and \eqref{qpnestedtwo}.

We continue the argument to show \eqref{recursionwavefunction}.
We note from the factor \eqref{factorinsummand} for each summand
in $E(\{ {\bm z^{(1)}} \}, \{ {\bm z^{(2)}} \}
|\{ {\bm w^{(1)}} \}, \{ {\bm w^{(2)}} \}
|\{k_1,k_2,L_1,L_2,{\bm I} \}|\lambda)$ that
only the summands satisfying $\sigma_2(k_2)=k_2$
in \eqref{multivariablefunction} survive after the substitution
$w_{L_2}^{(2)}=z_{k_2}^{(2)}-\gamma$.
Then we find that after the substitution
$w_{L_2}^{(2)}=z_{k_2}^{(2)}-\gamma$, the following
product of factors
\begin{align}
&[\gamma]^{k_1+k_2} \nonumber \\
\times& \prod_{a=1}^{k_2}
\Bigg(
\prod_{i=1}^{I_a^{(2)}-1} [z_{\sigma_2(a)}^{(2)}-w_i^{(2)}]
\frac{
[z_{\sigma_2(a)}^{(2)}-w_{I_a^{(2)}}^{(2)}+\lambda_3
-\lambda_{i_{I_a^{(2)}}^{(2)}}+(a-I_a^{(2)}-1+c(I_a^{(2)},i_{I_a^{(2)}}^{(2)}))\gamma]
}
{
[\lambda_{i_{I_a^{(2)}}^{(2)}}-\lambda_3
+(1-c(I_a^{(2)},i_{I_a^{(2)}}^{(2)}))\gamma]
}
\nonumber \\
\times& \prod_{i=I_a^{(2)}+1}^{L_2} [z_{\sigma_2(a)}^{(2)}-w_i^{(2)}-\gamma]
\Bigg)
\prod_{1 \le a < b \le k_2} 
\frac{[z_{\sigma_2(a)}^{(2)}-z_{\sigma_2(b)}^{(2)}+\gamma]}
{[z_{\sigma_2(a)}^{(2)}-z_{\sigma_2(b)}^{(2)}]},
\label{oneofthefactor}
\end{align}
in each summand in \eqref{multivariablefunction}
can be rewritten as
\begin{align}
&[\gamma]^{k_1+k_2-1} \nonumber \\
\times&\prod_{a=1}^{k_2-1} \Bigg(
\prod_{i=1}^{I_a^{(2)}-1} [z_{\sigma_2(a)}^{(2)}-w_i^{(2)}]
\frac{
[z_{\sigma_2(a)}^{(2)}-w_{I_a^{(2)}}^{(2)}+\lambda_3
-\lambda_{i_{I_a^{(2)}}^{(2)}}+(a-I_a^{(2)}-1+c(I_a^{(2)},i_{I_a^{(2)}}^{(2)}))\gamma]
}
{
[\lambda_{i_{I_a^{(2)}}^{(2)}}-\lambda_3
+(1-c(I_a^{(2)},i_{I_a^{(2)}}^{(2)}))\gamma]
}
\nonumber \\
\times& \prod_{i=I_a^{(2)}+1}^{L_2-1} [z_{\sigma_2(a)}^{(2)}-w_i^{(2)}-\gamma]
\Bigg)
\prod_{1 \le a < b \le k_2-1} 
\frac{[z_{\sigma_2(a)}^{(2)}-z_{\sigma_2(b)}^{(2)}+\gamma]}
{[z_{\sigma_2(a)}^{(2)}-z_{\sigma_2(b)}^{(2)}]} \nonumber \\
\times& \frac{[\gamma][\lambda_3-\lambda_{i^{(2)}_{L_2}}+\gamma(k_2-L_2+c(L_2,i^{(2)}_{L_2}))]}{[\lambda_{i_{L_2}^{(2)}}-\lambda_3+\gamma (1-c(L_2,i_{L_2}^{(2)}))]}
\prod_{j=1}^{k_2-1} [z_j^{(2)}-z_{k_2}^{(2)}+\gamma]
\prod_{j=1}^{L_2-1} [z_{k_2}^{(2)}-w_j^{(2)}].
\label{tochufactorone}
\end{align}
We further rewrite this using the set
${\bm J}=\{{\bm J_{k_1}^{(1)}}, {\bm J_{k_2-1}^{(2)}},
 \widehat{{\bm J}}_{(k_2-1)+(L_1+1)}^{(2)},
\widehat{{\bm J}}_{L_2-1}^{(3)} \}$
whose relation with the set
${\bm I}=\{{\bm I_{k_1}^{(1)}}, {\bm I_{k_2}^{(2)}},
 \widehat{{\bm I}}_{k_2+L_1}^{(2)}, \widehat{{\bm I}}_{L_2}^{(3)} \}$
is given in
Proposition \ref{higherrankpartitionfunctionsizerginkorepin}.
Since
${\bm J_{k_2-1}^{(2)}}
={\bm I_{k_2}^{(2)}} \backslash \{ L_2 \}$
and $I_{k_2}^{(2)}=L_2$,
we have the following relation
$J_a^{(2)}=I_a^{(2)}$, $a=1,\dots,k_2-1$.
Using this relation,
\eqref{tochufactorone} can be rewritten as
\begin{align}
&[\gamma]^{k_1+k_2-1} \nonumber \\
\times&\prod_{a=1}^{k_2-1} \Bigg(
\prod_{i=1}^{J_a^{(2)}-1} [z_{\sigma_2(a)}^{(2)}-w_i^{(2)}]
\frac{
[z_{\sigma_2(a)}^{(2)}-w_{J_a^{(2)}}^{(2)}+\lambda_3
-\lambda_{i_{J_a^{(2)}}^{(2)}}+(a-J_a^{(2)}-1+c(J_a^{(2)},i_{J_a^{(2)}}^{(2)}))\gamma]
}
{
[\lambda_{i_{J_a^{(2)}}^{(2)}}-\lambda_3
+(1-c(J_a^{(2)},i_{J_a^{(2)}}^{(2)}))\gamma]
}
\nonumber \\
\times& \prod_{i=J_a^{(2)}+1}^{L_2-1} [z_{\sigma_2(a)}^{(2)}-w_i^{(2)}-\gamma]
\Bigg)
\prod_{1 \le a < b \le k_2-1} 
\frac{[z_{\sigma_2(a)}^{(2)}-z_{\sigma_2(b)}^{(2)}+\gamma]}
{[z_{\sigma_2(a)}^{(2)}-z_{\sigma_2(b)}^{(2)}]} \nonumber \\
\times& \frac{[\gamma][\lambda_3-\lambda_{i^{(2)}_{L_2}}+\gamma(k_2-L_2+c(L_2,i^{(2)}_{L_2}))]}{[\lambda_{i_{L_2}^{(2)}}-\lambda_3+\gamma (1-c(L_2,i_{L_2}^{(2)}))]}
\prod_{j=1}^{k_2-1} [z_j^{(2)}-z_{k_2}^{(2)}+\gamma]
\prod_{j=1}^{L_2-1} [z_{k_2}^{(2)}-w_j^{(2)}].
\label{usethisexpressionone}
\end{align}

We next rewrite the remaining product of factors
\begin{align}
&\prod_{a=1}^{k_1}
\Bigg(
\prod_{i=1}^{\widetilde{I_a^{(1)}}-1}
[z_{\sigma_1(a)}^{(1)}-m_i^{k_2,L_1}
(\{ {\bm z_{\sigma_2}^{(2)}} \}| \{ {\bm w^{(1)}} \})
] \nonumber \\
\times&\frac{[z_{\sigma_1(a)}^{(1)}-m_{\widetilde{I_a^{(1)}}}^{k_2,L_1} (\{ {\bm z_{\sigma_2}^{(2)}} \}| \{ {\bm w^{(1)}} \})
+\lambda_2-\lambda_1+\gamma(2a-1-\widetilde{I_a^{(1)}})
]}
{[\lambda_1-\lambda_2+(1-a)\gamma]} \nonumber \\
\times&\prod_{i={\widetilde{I_a^{(1)}}+1}}^{k_2+L_1}
[z_{\sigma_1(a)}^{(1)}-m_i^{k_2,L_1}
(\{ {\bm z_{\sigma_2}^{(2)}} \}| \{ {\bm w^{(1)}} \})-\gamma
]
\Bigg) \prod_{1 \le a < b \le k_1} 
\frac{[z_{\sigma_1(a)}^{(1)}-z_{\sigma_1(b)}^{(1)}+\gamma]}
{[z_{\sigma_1(a)}^{(1)}-z_{\sigma_1(b)}^{(1)}]},
\label{tochufactortwo}
\end{align}
which, together with the factors \eqref{oneofthefactor},
forms each summand in \eqref{multivariablefunction}.
We again rewrite using the set
${\bm J}=\{{\bm J_{k_1}^{(1)}}, {\bm J_{k_2-1}^{(2)}},
 \widehat{{\bm J}}_{(k_2-1)+(L_1+1)}^{(2)},
\widehat{{\bm J}}_{L_2-1}^{(3)} \}$.

First, let us recall that only the summands satisfying $\sigma_2(k_2)=k_2$
in \eqref{multivariablefunction} survive after the substitution
$w_{L_2}^{(2)}=z_{k_2}^{(2)}-\gamma$.
When $\sigma_2(k_2)=k_2$, we can rewrite
$m_i^{k_2,L_1}(\{ {\bm z_{\sigma_2}^{(2)}} \}|\{ {\bm w^{(1)}} \})$
as
\begin{align}
&m_i^{k_2,L_1}(\{ {\bm z_{\sigma_2}^{(2)}} \}|\{ {\bm w^{(1)}} \})
= \begin{cases}
    z_{\sigma_2(i)}^{(2)} & 1 \le i \le k_2-1 \\
      z_{k_2}^{(2)}    &  i=k_2    \\
    w_{i-k_2}^{(1)} & k_2+1 \le i \le k_2+L_1
  \end{cases} \nonumber \\
=&m_i^{k_2-1,L_1+1}(\{ z_{\sigma_2(1)}^{(2)},\dots,
z_{\sigma_2(k_2-1)}^{(2)}
 \}|\{ 
z_{k_2}^{(2)}, w_1^{(1)}, \dots, w_{L_1}^{(1)}
\}),
\end{align}
which can be easily checked from the definition
of $m_i^{k_2,L_1}(\{ {\bm z^{(2)}} \}|\{ {\bm w^{(1)}} \})$
\eqref{definitionofm}.

We also note that since
${\bm J_{k_1}^{(1)}}={\bm I_{k_1}^{(1)}}$
and
$\widehat{{\bm J}}_{(k_2-1)+(L_1+1)}^{(2)}=
\widehat{{\bm I}}_{k_2+L_1}^{(2)}
$, the inclusion relation
$
{\bm J}_{k_1}^{(1)} \subset \widehat{{\bm J}}_{(k_2-1)+(L_1+1)}^{(2)}
$
for ${\bm J}$ is exactly the same with the one
$
{\bm I}_{k_1}^{(1)} \subset \widehat{{\bm I}}_{k_2+L_1}^{(2)}
$ for ${\bm I}$.
The induced sets $\widetilde{{\bm J_{k_1}^{(1)}}}$
and $\widetilde{{\bm I_{k_1}^{(1)}}}$ are induced from these inclusion
relations in the same way and since both relations are
exactly the same, we conclude
$\widetilde{{\bm J_{k_1}^{(1)}}}=\widetilde{{\bm I_{k_1}^{(1)}}}$.
Hence the elements of both sets
are all the same 
$\widetilde{J_a^{(1)}}=\widetilde{I_a^{(1)}}$, $a=1,\dots,k_1$
and one can rewrite \eqref{tochufactortwo}
using the set
${\bm J}=\{{\bm J_{k_1}^{(1)}}, {\bm J_{k_2-1}^{(2)}},
 \widehat{{\bm J}}_{(k_2-1)+(L_1+1)}^{(2)},
\widehat{{\bm J}}_{L_2-1}^{(3)} \}$ as

\begin{align}
&\prod_{a=1}^{k_1}
\Bigg(
\prod_{i=1}^{\widetilde{J_a^{(1)}}-1}
[z_{\sigma_1(a)}^{(1)}-
m_i^{k_2-1,L_1+1}(\{ z_{\sigma_2(1)}^{(2)},\dots,
z_{\sigma_2(k_2-1)}^{(2)}
 \}|\{ 
z_{k_2}^{(2)}, w_1^{(1)}, \dots, w_{L_1}^{(1)}
\})
] \nonumber \\
\times&\frac{[z_{\sigma_1(a)}^{(1)}-
m_{\widetilde{J_a^{(1)}}}^{k_2-1,L_1+1}
(\{ z_{\sigma_2(1)}^{(2)},\dots,
z_{\sigma_2(k_2-1)}^{(2)}
 \}|\{ 
z_{k_2}^{(2)}, w_1^{(1)}, \dots, w_{L_1}^{(1)}
\})
+\lambda_2-\lambda_1+\gamma(2a-1-\widetilde{J_a^{(1)}})
]}
{[\lambda_1-\lambda_2+(1-a)\gamma]} \nonumber \\
\times&\prod_{i={\widetilde{J_a^{(1)}}+1}}^{(k_2-1)+(L_1+1)}
[z_{\sigma_1(a)}^{(1)}-
m_i^{k_2-1,L_1+1}(\{ z_{\sigma_2(1)}^{(2)},\dots,
z_{\sigma_2(k_2-1)}^{(2)}
 \}|\{ 
z_{k_2}^{(2)}, w_1^{(1)}, \dots, w_{L_1}^{(1)}
\})
-\gamma
] \Bigg) \nonumber \\
\times&\prod_{1 \le a < b \le k_1} 
\frac{[z_{\sigma_1(a)}^{(1)}-z_{\sigma_1(b)}^{(1)}+\gamma]}
{[z_{\sigma_1(a)}^{(1)}-z_{\sigma_1(b)}^{(1)}]}.
\label{usethisexpressiontwo}
\end{align}
Combining the two factors
\eqref{usethisexpressionone} and \eqref{usethisexpressiontwo}
whose product gives each summand in the elliptic multivariable functions,
we find that
$E(\{ {\bm z^{(1)}} \}, \{ {\bm z^{(2)}} \}
|\{ {\bm w^{(1)}} \}, \{ {\bm w^{(2)}} \}
|\{k_1,k_2,L_1,L_2,{\bm I} \}|\lambda)$ evaluated at $w_{L_2}^{(2)}=z_{k_2}^{(2)}-\gamma$ can be expressed as
\begin{align}
&E(\{ {\bm z^{(1)}} \}, \{ {\bm z^{(2)}} \}
|\{ {\bm w^{(1)}} \}, \{ {\bm w^{(2)}} \}|\{k_1,k_2,L_1,L_2,{\bm I} \}|\lambda)
|_{w_{L_2}^{(2)}=z_{k_2}^{(2)}-\gamma}
\nonumber \\
=&[\gamma]^{k_1+k_2-1} \sum_{\sigma_1 \in S_{k_1}}
\sum_{\sigma_2 \in S_{k_2-1}} \nonumber \\
\times&\prod_{a=1}^{k_1}
\Bigg(
\prod_{i=1}^{\widetilde{J_a^{(1)}}-1}
[z_{\sigma_1(a)}^{(1)}-
m_i^{k_2-1,L_1+1}(\{ z_{\sigma_2(1)}^{(2)},\dots,
z_{\sigma_2(k_2-1)}^{(2)}
 \}|\{ 
z_{k_2}^{(2)}, w_1^{(1)}, \dots, w_{L_1}^{(1)}
\})
] \nonumber \\
\times&\frac{[z_{\sigma_1(a)}^{(1)}-
m_{\widetilde{J_a^{(1)}}}^{k_2-1,L_1+1}
(\{ z_{\sigma_2(1)}^{(2)},\dots,
z_{\sigma_2(k_2-1)}^{(2)}
 \}|\{ 
z_{k_2}^{(2)}, w_1^{(1)}, \dots, w_{L_1}^{(1)}
\})
+\lambda_2-\lambda_1+\gamma(2a-1-\widetilde{J_a^{(1)}})
]}
{[\lambda_1-\lambda_2+(1-a)\gamma]} \nonumber \\
\times&\prod_{i={\widetilde{J_a^{(1)}}+1}}^{(k_2-1)+(L_1+1)}
[z_{\sigma_1(a)}^{(1)}-
m_i^{k_2-1,L_1+1}(\{ z_{\sigma_2(1)}^{(2)},\dots,
z_{\sigma_2(k_2-1)}^{(2)}
 \}|\{ 
z_{k_2}^{(2)}, w_1^{(1)}, \dots, w_{L_1}^{(1)}
\})
-\gamma
] \Bigg) \nonumber \\
\times&\prod_{1 \le a < b \le k_1} 
\frac{[z_{\sigma_1(a)}^{(1)}-z_{\sigma_1(b)}^{(1)}+\gamma]}
{[z_{\sigma_1(a)}^{(1)}-z_{\sigma_1(b)}^{(1)}]} \nonumber \\
\times&\prod_{a=1}^{k_2-1} \Bigg(
\prod_{i=1}^{J_a^{(2)}-1} [z_{\sigma_2(a)}^{(2)}-w_i^{(2)}]
\frac{
[z_{\sigma_2(a)}^{(2)}-w_{J_a^{(2)}}^{(2)}+\lambda_3
-\lambda_{i_{J_a^{(2)}}^{(2)}}+(a-J_a^{(2)}-1+c(J_a^{(2)},i_{J_a^{(2)}}^{(2)}))\gamma]
}
{
[\lambda_{i_{J_a^{(2)}}^{(2)}}-\lambda_3
+(1-c(J_a^{(2)},i_{J_a^{(2)}}^{(2)}))\gamma]
}
\nonumber \\
\times& \prod_{i=J_a^{(2)}+1}^{L_2-1} [z_{\sigma_2(a)}^{(2)}-w_i^{(2)}-\gamma]
\Bigg)
\prod_{1 \le a < b \le k_2-1} 
\frac{[z_{\sigma_2(a)}^{(2)}-z_{\sigma_2(b)}^{(2)}+\gamma]}
{[z_{\sigma_2(a)}^{(2)}-z_{\sigma_2(b)}^{(2)}]} \nonumber \\
\times& \frac{[\gamma][\lambda_3-\lambda_{i^{(2)}_{L_2}}+\gamma(k_2-L_2+c(L_2,i^{(2)}_{L_2}))]}{[\lambda_{i_{L_2}^{(2)}}-\lambda_3+\gamma (1-c(L_2,i_{L_2}^{(2)}))]}
\prod_{j=1}^{k_2-1} [z_j^{(2)}-z_{k_2}^{(2)}+\gamma]
\prod_{j=1}^{L_2-1} [z_{k_2}^{(2)}-w_j^{(2)}] \nonumber \\
=&\frac{[\gamma][\lambda_3-\lambda_{i^{(2)}_{L_2}}+\gamma(k_2-L_2+c(L_2,i^{(2)}_{L_2}))]}{[\lambda_{i_{L_2}^{(2)}}-\lambda_3+\gamma (1-c(L_2,i_{L_2}^{(2)}))]}
\prod_{j=1}^{k_2-1} [z_j^{(2)}-z_{k_2}^{(2)}+\gamma]
\prod_{j=1}^{L_2-1} [z_{k_2}^{(2)}-w_j^{(2)}]
\nonumber \\
\times&
E(\{ {\bm z^{(1)}} \}, \{ {z_1^{(2)}, \dots, z_{k_2-1}^{(2)} } \}
|\{ z_{k_2}^{(2)}, w_1^{(1)}, \dots, w_{L_1}^{(1)} \}, \{ w_1^{(2)}, \dots,
w_{L_2-1}^{(2)} \} \nonumber \\
&|\{k_1,k_2-1,L_1+1,L_2-1,{\bm J} \}|\lambda).
\end{align}
This relation for the elliptic functions is exactly the same as the 
relation \eqref{recursionwavefunction}
for the partition functions
$W(\{ {\bm z^{(1)}} \}, \{ {\bm z^{(2)}} \}
|\{ {\bm w^{(1)}} \}, \{ {\bm w^{(2)}} \}|\{k_1,k_2,L_1,L_2,{\bm I} \}|\lambda)
$,
and hence property (3) for the case
$i_{L_2}^{(2)}=1$ or  $i_{L_2}^{(2)}=2$ is shown.

Let us show the case 
when $i_{L_2}^{(2)}=3$ which can be shown in a similar way.
We rewrite the elliptic functions using the set
${\bm K}=\{{\bm K_{k_1}^{(1)}}, {\bm K_{k_2}^{(2)}},
 \widehat{{\bm K}}_{k_2+L_1}^{(2)}, \widehat{{\bm K}}_{L_2-1}^{(3)} \}$,
whose relation with
${\bm I}:=\{{\bm I_{k_1}^{(1)}}, {\bm I_{k_2}^{(2)}},
 \widehat{{\bm I}}_{k_2+L_1}^{(2)}, \widehat{{\bm I}}_{L_2}^{(3)} \}$
is given in
Proposition \ref{higherrankpartitionfunctionsizerginkorepin}.

First, we note $K_{a}^{(2)}=I_a^{(2)}$, $a=1,\dots,k_2$
since ${\bm K_{k_2}^{(2)}}={\bm I_{k_2}^{(2)}}$.
Next, using ${\bm K}_{k_1}^{(1)}
={\bm I}_{k_1}^{(1)}|_{L_2+1 \rightarrow L_2,\dots,L_2+L_1 \rightarrow L_2+L_1-1}$ and $\widehat{{\bm K}}_{k_2+L_1}^{(2)}
={\bm K_{k_2}^{(2)}} \cup \{ L_2,\dots,L_2+L_1-1 \}
={\bm I_{k_2}^{(2)}} \cup \{ L_2,\dots,L_2+L_1-1 \}
={\bm I_{k_2}^{(2)}} \cup \{ L_2+1,\dots,L_2+L_1 \}|_{L_2+1 \rightarrow L_2,\dots,L_2+L_1 \rightarrow L_2+L_1-1}
=\widehat{{\bm I}}_{k_2+L_1}^{(2)}|_{L_2+1 \rightarrow L_2,\dots,L_2+L_1 \rightarrow L_2+L_1-1}
$,
one finds that the inclusion relation
${\bm K}_{k_1}^{(1)} \subset \widehat{{\bm K}}_{k_2+L_1}^{(2)}
:={\bm K_{k_2}^{(2)}} \cup \{ L_2,\dots,L_2+L_1-1 \}$
can be rewritten as
${\bm I}_{k_1}^{(1)}|_{L_2+1 \rightarrow L_2,\dots,L_2+L_1 \rightarrow L_2+L_1-1} \subset \widehat{{\bm I}}_{k_2+L_1}^{(2)}|_{L_2+1 \rightarrow L_2,\dots,L_2+L_1 \rightarrow L_2+L_1-1}$. From this rewriting, we find that
the induced sets induced by the mapping the inclusion relations
to $\{1,\dots,k_2+L_1 \}$
are exactly the same
$\widetilde{{\bm K_{k_1}^{(1)}}}=\widetilde{{\bm I_{k_1}^{(1)}}}$,
hence we get
$\widetilde{K_{a}^{(1)}}=\widetilde{I_a^{(1)}}$, $a=1,\dots,k_1$.
Finally, note that when $i_{L_2}^{(2)}=3$,
$I_{k_2}^{(2)} \le L_2-1$ holds since $L_2 \notin {\bm I_{k_2}^{(2)}}$,
and using this fact, one rewrites the factor
$\displaystyle \prod_{a=1}^{k_2} \prod_{i=I_a^{(2)}+1}^{L_2}
[z_{\sigma_2(a)}^{(2)}-w_i^{(2)}-\gamma]
$ in \eqref{multivariablefunction} as
$\displaystyle
\prod_{j=1}^{k_2} [z_j^{(2)}-w_{L_2}^{(2)}-\gamma]
\prod_{a=1}^{k_2} \prod_{i=I_a^{(2)}+1}^{L_2-1}
[z_{\sigma_2(a)}^{(2)}-w_i^{(2)}-\gamma]
$. Using this rewriting
and switching from the set
${\bm I}$ to ${\bm K}$ using the above rule, we find that
$E(\{ {\bm z^{(1)}} \}, \{ {\bm z^{(2)}} \}
|\{ {\bm w^{(1)}} \}, \{ {\bm w^{(2)}} \}
|\{k_1,k_2,L_1,L_2,{\bm I} \}|\lambda)$ can be expressed as

\begin{align}
&E(\{ {\bm z^{(1)}} \}, \{ {\bm z^{(2)}} \}
|\{ {\bm w^{(1)}} \}, \{ {\bm w^{(2)}} \}|\{k_1,k_2,L_1,L_2,{\bm I} \}|\lambda)
\nonumber \\
=&
\Bigg(\prod_{j=1}^{k_2} [z_j^{(2)}-w_{L_2}^{(2)}-\gamma] \Bigg)
[\gamma]^{k_1+k_2}
\sum_{\sigma_1 \in S_{k_1}} \sum_{\sigma_2 \in S_{k_2}}
\prod_{a=1}^{k_1}
\Bigg(
\prod_{i=1}^{\widetilde{K_a^{(1)}}-1}
[z_{\sigma_1(a)}^{(1)}-m_i^{k_2,L_1}
(\{ {\bm z_{\sigma_2}^{(2)}} \}| \{ {\bm w^{(1)}} \})
] \nonumber \\
\times&
\frac{[z_{\sigma_1(a)}^{(1)}-m_{\widetilde{K_a^{(1)}}}^{k_2,L_1}
(\{ {\bm z_{\sigma_2}^{(2)}} \}| \{ {\bm w^{(1)}} \})
+\lambda_2-\lambda_1+\gamma(2a-1-\widetilde{K_a^{(1)}})
]}
{[\lambda_1-\lambda_2+(1-a)\gamma]} \nonumber \\
\times&\prod_{i={\widetilde{K_a^{(1)}}+1}}^{k_2+L_1}
[z_{\sigma_1(a)}^{(1)}-m_i^{k_2,L_1}
(\{ {\bm z_{\sigma_2}^{(2)}} \}| \{ {\bm w^{(1)}} \})-\gamma
]
\Bigg) \nonumber \\
\times&\prod_{1 \le a < b \le k_1} 
\frac{[z_{\sigma_1(a)}^{(1)}-z_{\sigma_1(b)}^{(1)}+\gamma]}
{[z_{\sigma_1(a)}^{(1)}-z_{\sigma_1(b)}^{(1)}]}
\prod_{a=1}^{k_2} \Bigg(
\prod_{i=1}^{K_a^{(2)}-1} [z_{\sigma_2(a)}^{(2)}-w_i^{(2)}]
\nonumber \\
\times&\frac{
[z_{\sigma_2(a)}^{(2)}-w_{K_a^{(2)}}^{(2)}+\lambda_3
-\lambda_{i_{K_a^{(2)}}^{(2)}}+(a-K_a^{(2)}-1+c(K_a^{(2)},i_{K_a^{(2)}}^{(2)}))\gamma]
}
{
[\lambda_{i_{K_a^{(2)}}^{(2)}}-\lambda_3
+(1-c(K_a^{(2)},i_{K_a^{(2)}}^{(2)}))\gamma]
}
\nonumber \\
\times&\prod_{i=K_a^{(2)}+1}^{L_2-1} [z_{\sigma_2(a)}^{(2)}-w_i^{(2)}-\gamma]
\Bigg)
\prod_{1 \le a < b \le k_2} 
\frac{[z_{\sigma_2(a)}^{(2)}-z_{\sigma_2(b)}^{(2)}+\gamma]}
{[z_{\sigma_2(a)}^{(2)}-z_{\sigma_2(b)}^{(2)}]} \nonumber \\
=&
\prod_{j=1}^{k_2} [z_j^{(2)}-w_{L_2}^{(2)}-\gamma] \nonumber \\
\times&
E(\{ {\bm z^{(1)}} \}, \{ {\bm z^{(2)}} \}
|\{ {\bm w^{(1)}} \}, \{ w_1^{(2)},\dots, w_{L_2-1}^{(2)} \}|\{k_1,k_2,L_1,L_2-1,{\bm K} \}|\lambda),
\end{align}
hence we have shown that the elliptic functions satisfy
the factorization property \eqref{recursionwavefunction2}.

Property (4) for the case $k_2=1$ and $i^{(2)}_{L_2}=1$ or $i^{(2)}_{L_2}=2$
can be checked as follows.
In this case, we have $I_1^{(2)}=L_2$ and $c(L_2,i_{L_2}^{(2)})=1$
since $i_1^{(2)}= \cdots =i_{L_2-1}^{(2)}=3$.
One can also easily see that the elements of the induced set
$\widetilde{{\bm I_{k_1}^{(1)}}}$
are given by those of the set ${\bm I_{k_1}^{(1)}}$
as $\widetilde{I_a^{(1)}}=I_a^{(1)}+1-L_2$, $a=1,\dots,k_2$.
Then one finds that
\eqref{multivariablefunction}
can be rewritten as

\begin{align}
&E(\{ {\bm z^{(1)}} \}, \{ {z_1^{(2)}} \}
|\{ {\bm w^{(1)}} \}, \{ {\bm w^{(2)}} \}|\{k_1,1,L_1,L_2,{\bm I} \}|\lambda)
\nonumber \\
=&[\gamma]^{k_1+1}
\sum_{\sigma_1 \in S_{k_1}}
\prod_{a=1}^{k_1}
\Bigg(
\prod_{i=1}^{\widetilde{I_a^{(1)}}-1}
[z_{\sigma_1(a)}^{(1)}-m_i^{1,L_1}
(\{ z_{1}^{(2)} \}| \{ {\bm w^{(1)}} \})
] \nonumber \\
\times&
\frac{[z_{\sigma_1(a)}^{(1)}-m_{\widetilde{I_a^{(1)}}}^{1,L_1}
(\{ z_{1}^{(2)} \}| \{ {\bm w^{(1)}} \})
+\lambda_2-\lambda_1+\gamma(2a-1-\widetilde{I_a^{(1)}})
]}
{[\lambda_1-\lambda_2+(1-a)\gamma]} \nonumber \\
\times&\prod_{i={\widetilde{I_a^{(1)}}+1}}^{L_1+1}
[z_{\sigma_1(a)}^{(1)}-m_i^{1,L_1}
(\{  z_{1}^{(2)} \}| \{ {\bm w^{(1)}} \})-\gamma
]
\Bigg) \prod_{1 \le a < b \le k_1} 
\frac{[z_{\sigma_1(a)}^{(1)}-z_{\sigma_1(b)}^{(1)}+\gamma]}
{[z_{\sigma_1(a)}^{(1)}-z_{\sigma_1(b)}^{(1)}]}
\nonumber \\
\times&\prod_{i=1}^{L_2-1} [z_1^{(2)}-w_i^{(2)}]
\frac{
[z_{1}^{(2)}-w_{L_2}^{(2)}+\lambda_3
-\lambda_{i_{L_2}^{(2)}}+(1-L_2)\gamma]
}
{
[\lambda_{i_{L_2}^{(2)}}-\lambda_3]
} \nonumber \\
=&[\gamma]
\prod_{i=1}^{L_2-1} [z_1^{(2)}-w_i^{(2)}]
\frac{
[z_{1}^{(2)}-w_{L_2}^{(2)}+\lambda_3
-\lambda_{i_{L_2}^{(2)}}+(1-L_2)\gamma]
}
{
[\lambda_{i_{L_2}^{(2)}}-\lambda_3]
} \nonumber \\
\times&[\gamma]^{k_1}
\sum_{\sigma_1 \in S_{k_1}}
\prod_{a=1}^{k_1}
\Bigg(
\prod_{i=1}^{\widetilde{I_a^{(1)}}-1}
[z_{\sigma_1(a)}^{(1)}-m_i^{1,L_1}
(\{ z_{1}^{(2)} \}| \{ {\bm w^{(1)}} \})
] \nonumber \\
\times&
\frac{[z_{\sigma_1(a)}^{(1)}-
m_{\widetilde{I_a^{(1)}}}^{1,L_1}
(\{ z_{1}^{(2)} \}| \{ {\bm w^{(1)}} \})
+\lambda_2-\lambda_1+\gamma(2a-1-\widetilde{I_a^{(1)}})
]}
{[\lambda_1-\lambda_2+(1-a)\gamma]}
\nonumber \\
\times&\prod_{i={\widetilde{I_a^{(1)}}+1}}^{L_1+1}
[z_{\sigma_1(a)}^{(1)}-m_i^{1,L_1}
(\{  z_{1}^{(2)} \}| \{ {\bm w^{(1)}} \})-\gamma
]
\Bigg) \prod_{1 \le a < b \le k_1} 
\frac{[z_{\sigma_1(a)}^{(1)}-z_{\sigma_1(b)}^{(1)}+\gamma]}
{[z_{\sigma_1(a)}^{(1)}-z_{\sigma_1(b)}^{(1)}]}
\nonumber \\
=&
\frac{[\gamma][z_1^{(2)}-w_{L_2}^{(2)}+\lambda_3-\lambda_{i_{L_2}^{(2)}}-\gamma(L_2-1)]}{[\lambda_{i_{L_2}^{(2)}}-\lambda_3]}
\prod_{j=1}^{L_2-1} [z_1^{(2)}-w_j^{(2)}] \nonumber \\
\times&
E_{L_1+1,k_1}(z_1^{(1)},\dots,z_{k_1}^{(1)}|
z_1^{(2)},w_1^{(1)},\dots,w_{L_1}^{(1)}|\widetilde{I_1^{(1)}},\dots,\widetilde{I_{k_1}^{(1)}}|\lambda)
\nonumber \\
=&
\frac{[\gamma][z_1^{(2)}-w_{L_2}^{(2)}+\lambda_3-\lambda_{i_{L_2}^{(2)}}-\gamma(L_2-1)]}{[\lambda_{i_{L_2}^{(2)}}-\lambda_3]}
\prod_{j=1}^{L_2-1} [z_1^{(2)}-w_j^{(2)}]
\nonumber \\
\times&
E_{L_1+1,k_1}(z_1^{(1)},\dots,z_{k_1}^{(1)}|
z_1^{(2)},w_1^{(1)},\dots,w_{L_1}^{(1)}|I_1^{(1)}+1-L_2,\dots,I_{k_1}^{(1)}+1-L_2|\lambda),
\end{align}
hence one concludes that the elliptic multivariable functions
satisfy the initial conditions
\eqref{initialrecursion}.

Since we have shown that the functions
$E(\{ {\bm z^{(1)}} \}, \{ {\bm z^{(2)}} \}
|\{ {\bm w^{(1)}} \}, \{ {\bm w^{(2)}} \}
|\{k_1,k_2,L_1,L_2,{\bm I} \}|\lambda)$ satisfy all the
properties in
Proposition \ref{higherrankpartitionfunctionsizerginkorepin},
they are the explicit forms of the
partition functions of Foda-Manabe type
\begin{align}
&W(\{ {\bm z^{(1)}} \}, \{ {\bm z^{(2)}} \}
|\{ {\bm w^{(1)}} \}, \{ {\bm w^{(2)}} \}
|\{k_1,k_2,L_1,L_2,{\bm I} \}|\lambda) \nonumber \\
=&
E(\{ {\bm z^{(1)}} \}, \{ {\bm z^{(2)}} \}
|\{ {\bm w^{(1)}} \}, \{ {\bm w^{(2)}} \}
|\{k_1,k_2,L_1,L_2,{\bm I} \}|\lambda). \nonumber
\end{align}

\end{proof}

\section{Connections with Elliptic weight functions}
In this section, we show that special cases of
the elliptic partition functions of Foda-Manabe type introduced
in this paper are essentially the elliptic weights functions introduced in
the works by Rim\'anyi-Tarasov-Varchenko, Konno, Felder-Rim\'anyi-Varchenko
\cite{Konno1,Konno2,FRV,RTV2}.
The elliptic weight functions appear in the integral representation of the solutions to the elliptic $q$-KZ equation,
which is a difference equation described by the elliptic dynamical $R$-matrix.
The elliptic weight functions also become
the elliptic stable envelopes for the cotangent bundles of flag varieties
by an appropriate projection procedure,
which are meromorphic sections of certain line bundles
on products of elliptic curves,
and define the equivariant elliptic cohomology class
originally proposed by Aganagic-Okounkov \cite{AO}, which is
an elliptic generalization of the work by Maulik-Okounkov \cite{MO}
in which they
initiated a program to relate quantum torus equivariant cohomology
of quiver varieties and representation theory of quantum groups.
See \cite{Konno1,Konno2,FRV,RTV2} for details about the
elliptic weight functions and elliptic stable envelopes.

The elliptic weight functions was introduced as an elliptic analogue
of the trigonometric weight functions, and it was shown that
the trigonometric weight functions have connections with the
off-shell Bethe wavefunctions of trigonometric integrable models \cite{TV}.
From this fact, it is supposed that the special case $L_1=0$
of the elliptic multivariable functions \eqref{multivariablefunction}
introduced in this paper
have connections with the elliptic weight functions,
since the type of partition functions of the
off-shell Bethe wavefunctions of the trigonometric $U_q(\widehat{sl_3})$ model
correspond to the case $L_1=0$ \cite{TarVarSigma}.
The same type of partition functions using rational $SU(3)$ $R$-matrix
first appeared in \cite{Reshetikhin}.
See \cite{BW} for applications of this type of partition functions
to  probability theory.
See also \cite{PRS,Ros} for example for the 
relation between the domain wall boundary partition functions 
constructed from the elliptic dynamical $R$-matrix 
$E_{\tau,\eta}(gl_2)$ and the simplest case of the weight functions.

In this section, we consider the case $L_1=0$ of the elliptic multivariable functions \eqref{multivariablefunction}.
When $L_1=0$, first note that 
$m_i^{k_2,L_1}(\{ {\bm z^{(2)}} \}|\{ {\bm w^{(1)}} \})$
defined as \eqref{definitionofm} becomes
$m_i^{k_2,L_1}(\{ {\bm z^{(2)}} \}|\{ {\bm w^{(1)}} \}=
z_i^{(2)}, \ 1 \le i \le k_2$. Then we find
the elliptic multivariable functions
\eqref{multivariablefunction} can be written as
\begin{align}
&E(\{ {\bm z^{(1)}} \}, \{ {\bm z^{(2)}} \}
|\phi, \{ {\bm w^{(2)}} \}|\{k_1,k_2,0,L_2,{\bm I} \}|\lambda)
\nonumber \\
=&[\gamma]^{k_1+k_2}
\sum_{\sigma_1 \in S_{k_1}} \sum_{\sigma_2 \in S_{k_2}}
\prod_{a=1}^{k_1}
\Bigg(
\prod_{i=1}^{\widetilde{I_a^{(1)}}-1}
[z_{\sigma_1(a)}^{(1)}-z_{\sigma_2(i)}^{(2)}
] \nonumber \\
\times&
\frac{[z_{\sigma_1(a)}^{(1)}-
z_{\sigma_2(\widetilde{I_a^{(1)}})}^{(2)}
+\lambda_2-\lambda_1+\gamma(2a-1-\widetilde{I_a^{(1)}})
]}
{[\lambda_1-\lambda_2+(1-a)\gamma]}
\prod_{i={\widetilde{I_a^{(1)}}+1}}^{k_2}
[z_{\sigma_1(a)}^{(1)}-z_{\sigma_2(i)}^{(2)}-\gamma
]
\Bigg) \nonumber \\
\times&\prod_{1 \le a < b \le k_1} 
\frac{[z_{\sigma_1(a)}^{(1)}-z_{\sigma_1(b)}^{(1)}+\gamma]}
{[z_{\sigma_1(a)}^{(1)}-z_{\sigma_1(b)}^{(1)}]}
\prod_{a=1}^{k_2} \Bigg(
\prod_{i=1}^{I_a^{(2)}-1} [z_{\sigma_2(a)}^{(2)}-w_i^{(2)}]
\nonumber \\
\times&\frac{
[z_{\sigma_2(a)}^{(2)}-w_{I_a^{(2)}}^{(2)}+\lambda_3
-\lambda_{i_{I_a^{(2)}}^{(2)}}+(a-I_a^{(2)}-1+c(I_a^{(2)},i_{I_a^{(2)}}^{(2)}))\gamma]
}
{
[\lambda_{i_{I_a^{(2)}}^{(2)}}-\lambda_3
+(1-c(I_a^{(2)},i_{I_a^{(2)}}^{(2)}))\gamma]
}
\nonumber \\
\times&\prod_{i=I_a^{(2)}+1}^{L_2} [z_{\sigma_2(a)}^{(2)}-w_i^{(2)}-\gamma]
\Bigg)
\prod_{1 \le a < b \le k_2} 
\frac{[z_{\sigma_2(a)}^{(2)}-z_{\sigma_2(b)}^{(2)}+\gamma]}
{[z_{\sigma_2(a)}^{(2)}-z_{\sigma_2(b)}^{(2)}]}.
\label{comparisonpre}
\end{align}

In order to make comparison
with the elliptic weight functions clearer,
we multiply the elliptic multivariable functions \eqref{comparisonpre}
by the following overall factor which is independent
of the spectral parameters
\begin{align}
[\gamma]^{-k_1-k_2}
\prod_{a=1}^{k_1}[\lambda_1-\lambda_2+(1-a)\gamma]
\prod_{a=1}^{k_2}
[\lambda_{i_{I_a^{(2)}}^{(2)}}-\lambda_3
+(1-c(I_a^{(2)},i_{I_a^{(2)}}^{(2)}))\gamma],
\end{align}
i.e., clear the overall factor in front of the double summation in
\eqref{comparisonpre} and the denominators which are independent of the
spectral parameters.
We further shift the spectral parameters $z_j^{(1)}$, $j=1,\dots,k_1$
by $+\gamma$, and shift $w_j^{(2)}$, $j=1,\dots,k_2$ by $-\gamma$.
These shifts correspond to the redefinition of the spectral parameters
$z_j^{(1)}$ and $w_j^{(2)}$.
We denote the elliptic multivariable functions
obtained by this modification as
$\overline{E}(\{ {\bm z^{(1)}} \}, \{ {\bm z^{(2)}} \}
| \phi , \{ {\bm w^{(2)}} \}|\{k_1,k_2,0,L_2,{\bm I} \}|\lambda)$,
which has the following representation
\begin{align}
&\overline{E}(\{ {\bm z^{(1)}} \}, \{ {\bm z^{(2)}} \}
| \phi , \{ {\bm w^{(2)}} \}|\{k_1,k_2,0,L_2,{\bm I} \}|\lambda)
\nonumber \\
=&
\sum_{\sigma_1 \in S_{k_1}} \sum_{\sigma_2 \in S_{k_2}}
\prod_{a=1}^{k_1}
\Bigg(
\prod_{i=1}^{\widetilde{I_a^{(1)}}-1}
[z_{\sigma_1(a)}^{(1)}-z_{\sigma_2(i)}^{(2)}+\gamma
] \nonumber \\
\times&
[z_{\sigma_1(a)}^{(1)}-
z_{\sigma_2(\widetilde{I_a^{(1)}})}^{(2)}
+\lambda_2-\lambda_1+\gamma(2a-\widetilde{I_a^{(1)}})
]
\prod_{i={\widetilde{I_a^{(1)}}+1}}^{k_2}
[z_{\sigma_1(a)}^{(1)}-z_{\sigma_2(i)}^{(2)}]
\Bigg) \nonumber \\
\times&\prod_{1 \le a < b \le k_1} 
\frac{[z_{\sigma_1(a)}^{(1)}-z_{\sigma_1(b)}^{(1)}+\gamma]}
{[z_{\sigma_1(a)}^{(1)}-z_{\sigma_1(b)}^{(1)}]}
\prod_{a=1}^{k_2} \Bigg(
\prod_{i=1}^{I_a^{(2)}-1} [z_{\sigma_2(a)}^{(2)}-w_i^{(2)}+\gamma]
\nonumber \\
\times&
[z_{\sigma_2(a)}^{(2)}-w_{I_a^{(2)}}^{(2)}+\lambda_3
-\lambda_{i_{I_a^{(2)}}^{(2)}}+(a-I_a^{(2)}+c(I_a^{(2)},i_{I_a^{(2)}}^{(2)}))\gamma]
\nonumber \\
\times&\prod_{i=I_a^{(2)}+1}^{L_2} [z_{\sigma_2(a)}^{(2)}-w_i^{(2)}]
\Bigg)
\prod_{1 \le a < b \le k_2} 
\frac{[z_{\sigma_2(a)}^{(2)}-z_{\sigma_2(b)}^{(2)}+\gamma]}
{[z_{\sigma_2(a)}^{(2)}-z_{\sigma_2(b)}^{(2)}]}. \label{forcomparison}
\end{align}

Now we review the explicit forms of the elliptic weight functions
introduced in \cite{Konno1,Konno2,FRV,RTV2}.
We keep the notations used in \cite{RTV2} for the description
of the elliptic weight functions.

The elliptic weight functions are introduced
using the theta function in the following multiplicative form
\begin{align}
\vartheta(x)=(x^{1/2}-x^{-1/2})\phi(qx)\phi(q/x),
\ \ \ \phi(x)=\prod_{s=0}^\infty (1-q^s x).
\end{align}
The theta function in the additive form \eqref{theta}
can be obtained from $\vartheta(x)$ by substituting
$q=e^{2 \pi i \tau}, x=e^{2 \pi i z}$.

The following notations are used in \cite{RTV2}
for the sets of indices and variables.
Fix two integers $n,N \in \mathbb{Z}_{\ge 0}$ and let ${\bm \lambda} \in \mathbb{Z}_{\ge 0}^N$ be such that $\sum_{k=1}^N \lambda_k=n$.
Set $\lambda^{(k)}=\lambda_1+\cdots+\lambda_k$ for $k=0,\dots,N$,
and $\lambda^{ \{ 1 \} }=\lambda^{(1)}+\cdots+\lambda^{(N-1)}$.
We define $\mathcal{I}_{\mathbb{\lambda}}$ as the set of partitions
$I=(I_1,\dots,I_N)$ of $\{1,\dots,n \}$ with $|I_k|=\lambda_k$.
For $I \in \mathcal{I}_{\mathbb{\lambda}}$,
the following notation
\begin{align}
\bigcup_{a=1}^k I_a=\{ i_1^{(k)}
< \cdots < i_{\lambda^{(k)}}^{(k)} \}, \label{followingnotation}
\end{align}
is used.

Introduce a variable $h$ and sets of variables ${\bm z}=\{z_1,\dots,z_n \}$,
${\bm \mu}=\{\mu_1,\dots,\mu_N \}$.
We also consider variables $t_a^{(k)}$ for $k=1,\dots,N$, $a=1,\dots,\lambda^{(k)}$, where $t_a^{(N)}=z_a$, $a=1,\dots,n$,
and denote $t^{(j)}=(t_k^{(j)})_{k \le \lambda^{(j)}}$
and ${\bm t}=(t^{(1)},\dots,t^{(N-1)})$.

For $I \in \mathcal{I}_{{\bm \lambda}}$, the elliptic weight function
is defined as \cite{Konno1,Konno2,FRV,RTV2}
\begin{align}
W_{I}^{\mathrm{ell}}({\bm t},{\bm z},h,{\bm \mu})
=( \vartheta(h) )^{\lambda^{ \{ 1 \} }}
\mathrm{Sym}_{t^{(1)}} \dots \mathrm{Sym}_{t^{(N-1)}}
U_I^{\mathrm{ell}}({\bm t}, {\bm z},h, {\bm \mu}),
\label{ellipticweightfunctions}
\end{align}
where $\mathrm{Sym}_{t^{(k)}}$ is the symmetrization with respect to the
variables $t_1^{(k)},\dots,t_{\lambda^{(k)}}^{(k)}$,
\begin{align}
\mathrm{Sym}_{t^{(k)}} f(t_1^{(k)},\dots,t_{\lambda^{(k)}}^{(k)})
=\sum_{\sigma \in S_{\lambda^{(k)}}}
f(t_{\sigma(1)}^{(k)},\dots,t_{\sigma(\lambda^{(k)})}^{(k)}),
\end{align}
and
\begin{align}
U_I^{\mathrm{ell}}({\bm t}, {\bm z},h, {\bm \mu})=
\prod_{k=1}^{N-1} \prod_{a=1}^{\lambda^{(k)}}
\Bigg(
\prod_{c=1}^{\lambda^{(k+1)}}
\psi_{I,k,a,c}^{\mathrm{ell}}(t_c^{(k+1)}/t_a^{(k)})
\prod_{b=a+1}^{\lambda^{(k)}} \frac{\vartheta(ht_b^{(k)}/t_a^{(k)})}
{\vartheta(t_b^{(k)}/t_a^{(k)})}
\Bigg), \label{component}
\end{align}
where
\begin{align}
\psi_{I,k,a,c}^{\mathrm{ell}}(x)
=
\begin{cases}
    \vartheta(hx), & i_c^{(k+1)}<i_a^{(k)}, \\
   \frac{\displaystyle \vartheta(
x h^{1+p_{I,j(I,k,a)}(i_a^{(k)})-p_{I,k+1}(i_a^{(k)})} \mu_{k+1}/\mu_{j(I,k,a)}
)}
{\displaystyle \vartheta(
h^{1+p_{I,j(I,k,a)}(i_a^{(k)})-p_{I,k+1}(i_a^{(k)})} \mu_{k+1}/\mu_{j(I,k,a)}
)},  & i_c^{(k+1)}=i_a^{(k)}, \\
     \vartheta(x),  & i_c^{(k+1)}>i_a^{(k)},
\end{cases}.
\label{cases}
\end{align}
Here, $j(I,k,a) \in \{1,\dots,N \}$ is such that
$i_a^{(k)} \in I_{j(I,k,a)}$ and
\begin{align}
p_{I,j}(m)=|I_j \cap \{1,\dots,m-1 \}|, \ \ \ j=1,\dots,N.
\end{align}
The product of all denominators which appear in \eqref{component}
and only depends on $h$ and ${\bm \mu}=\{\mu_1,\dots,\mu_N \}$ only
is denoted as
\begin{align}
\psi_{I}(h,{\bm \mu})
=\prod_{k=1}^{N-1} \prod_{k=1}^{\lambda^{(k)}}
\vartheta(
h^{1+p_{I,j(I,k,a)}(i_a^{(k)})-p_{I,k+1}(i_a^{(k)})} \mu_{k+1}/\mu_{j(I,k,a)}
).
\end{align}
To compare with \eqref{forcomparison},
we consider the elliptic weight functions
\eqref{ellipticweightfunctions} multiplied by
the factor $( \vartheta(h) )^{-\lambda^{ \{ 1 \} }} \psi_{I}(h,{\bm \mu})$, which we denote as $\overline{W}_{I}^{\mathrm{ell}}({\bm t},{\bm z},h,{\bm \mu})$
\begin{align}
\overline{W}_{I}^{\mathrm{ell}}({\bm t},{\bm z},h,{\bm \mu})
=( \vartheta(h) )^{-\lambda^{ \{ 1 \} }} \psi_{I}(h,{\bm \mu})
W_{I}^{\mathrm{ell}}({\bm t},{\bm z},h,{\bm \mu}).
\label{weightforcomparison}
\end{align}

Now we compare the case $N=3$ of the elliptic weight functions
$\overline{W}_{I}^{\mathrm{ell}}({\bm t},{\bm z},h,{\bm \mu})$
\eqref{weightforcomparison}
and 
$\overline{E}(\{ {\bm z^{(1)}} \}, \{ {\bm z^{(2)}} \}
| \phi , \{ {\bm w^{(2)}} \}|\{k_1,k_2,0,L_2,{\bm I} \}|\lambda)
$ \eqref{forcomparison}.

First, we identify the sets for configurations.
Recall the Foda-Manabe label introduced
in section 3 for the configurations used
in \eqref{forcomparison}.
The following set 
${\bm I}:=\{{\bm I_{k_1}^{(1)}}, {\bm I_{k_2}^{(2)}},
 \widehat{{\bm I}}_{k_2+L_1}^{(2)}, \widehat{{\bm I}}_{L_2}^{(3)} \}$
where the subsets
${\bm I_{k_1}^{(1)}}, {\bm I_{k_2}^{(2)}},
 \widehat{{\bm I}}_{k_2+L_1}^{(2)}, \widehat{{\bm I}}_{L_2}^{(3)} $
satisfy the following relations
\begin{align}
&{\bm I}_{k_1}^{(1)} \subset \widehat{{\bm I}}_{k_2+L_1}^{(2)}
:={\bm I_{k_2}^{(2)}} \cup \{ L_2+1,\dots,L_2+L_1 \}, \\
&{\bm I_{k_2}^{(2)}} \subset \widehat{{\bm I}}_{L_2}^{(3)}:=\{1,\dots,L_2 \},
\end{align}
is used for the label of a configuration of colors.
Since we are considering the case $L_1=0$,
$\widehat{{\bm I}}_{k_2+L_1}^{(2)}={\bm I}_{k_2}^{(2)}$ holds
and we can see that the three sets
${\bm I}_{k_1}^{(1)}, {\bm I}_{k_2}^{(2)}, \widehat{{\bm I}}_{L_2}^{(3)} $
contain the information of the configurations of colors.
${\bm I}_{k_1}^{(1)}$ is the set of positions which have color $1$,
and $\widehat{{\bm I}}_{k_2+L_1}^{(2)} \backslash {\bm I}_{k_1}^{(1)}$,
which is ${\bm I}_{k_2}^{(2)} \backslash {\bm I}_{k_1}^{(1)}$ for the
case we consider, records the positions of color $2$.
$\widehat{{\bm I}}_{L_2}^{(3)} \backslash {\bm I_{k_2}^{(2)}}
=\{1,\dots,L_2 \} \backslash {\bm I_{k_2}^{(2)}}$
is the set of positions which have color $3$.
Keeping these meanings for the sets in mind and
comparing with the sets \eqref{followingnotation}
used for the description of the elliptic weight functions,
we find it is natural to have the following identifications
between the labels for sets and numbers used in \cite{RTV2} and in this paper.
\begin{align}
&\bigcup_{a=1}^1 I_a \longleftrightarrow   {\bm I}_{k_1}^{(1)} , \ \ \
\bigcup_{a=1}^2 I_a \longleftrightarrow   {\bm I}_{k_2}^{(2)} , \ \ \
\bigcup_{a=1}^3 I_a \longleftrightarrow   \widehat{{\bm I}}_{L_2}^{(3)}, 
\label{preuseformatch1}  \\
&\lambda^{(1)} \longleftrightarrow k_1, \ \ \
\lambda^{(2)} \longleftrightarrow k_2, \ \ \
\lambda^{(3)}=n \longleftrightarrow L_2. \label{preuseformatch2}
\end{align}

We further examine the correspondence between the labels
used in \cite{RTV2} and in this paper.
$p_{I,j(I,2,a)}(i_a^{(2)})$ in \cite{RTV2}
counts the number of positions
from the 1st to the $(i_a^{(2)}-1)$th positions
which are colored with the same color in
the $i_a^{(2)}$th position.
$i_a^{(2)}$ in \cite{RTV2} corresponds to $I_a^{(2)}$ in this paper
since $\bigcup_{a=1}^2 I_a=\{i_1^{(2)} < \cdots < i_{\lambda^{(2)}}^{(2)} \}$
and ${\bm I}_{k_2}^{(2)}=\{I_1^{(2)}< \cdots < I_{k_2}^{(2)} \}$
are both the sets which label the positions colored
either by color 1 or color 2. In the notation used in this paper, this means that $p_{I,j(I,2,a)}(i_a^{(2)})$ corresponds
to
$
\# \{\ell | 1 \le \ell \le I_a^{(2)}-1, i_\ell^{(2)}=i_{I_a^{(2)}}^{(2)} \}
=\# \{\ell | 1 \le \ell \le I_a^{(2)}, i_\ell^{(2)}=i_{I_a^{(2)}}^{(2)} \}-1
=c(I_a^{(2)},i_{I_a^{(2)}}^{(2)})-1,
$
where we used the notation for the counting \eqref{countingoperator}.

Next, we see that $p_{I,3}(i_a^{(2)})$ in \cite{RTV2}
counts the number of positions
from the 1st to the $(i_a^{(2)}-1)$th positions
which are colored by color 3.
Since $I_a^{(2)}$ used in this paper labels the position of the $a$th
place which is colored either by color 1 or color 2 (not by color 3),
we can see that $p_{I,3}(i_a^{(2)})$ corresponds to $I_a^{(2)}-a$.
Combining this correspondence with the one between
$p_{I,j(I,2,a)}(i_a^{(2)})$ and $c(I_a^{(2)},i_{I_a^{(2)}}^{(2)})-1$,
we have the following correspondence
\begin{align}
p_{I,j(I,2,a)}(i_a^{(2)})-p_{I,3}(i_a^{(2)})
\longleftrightarrow
c(I_a^{(2)},i_{I_a^{(2)}}^{(2)})-1-I_a^{(2)}+a. \label{useformatch1}
\end{align}
Similar considerations for the correspondence of counting things
between \cite{RTV2} and in this paper can be done, and one
finds the correspondence
\begin{align}
p_{I,j(I,1,a)}(i_a^{(1)})-p_{I,2}(i_a^{(1)})
\longleftrightarrow
2a-1-\widetilde{I_a^{(1)}}. \label{useformatch2}
\end{align}
In the description of the double product of the elliptic
multivariable functions in \cite{RTV2},
the (in)equalities $i_c^{(2)}=i_a^{(1)}$,
$i_c^{(2)}>i_a^{(1)}$ and $i_c^{(2)}<i_a^{(1)}$ are used
in \eqref{cases}.
We can see that $i_c^{(2)}=i_a^{(1)}$ in \cite{RTV2}
corresponds to $c=\widetilde{I_a^{(1)}}$ in this paper,
since the equality $i_c^{(2)}=i_a^{(1)}$
means that $I_c^{(2)}$, which
labels the position of the $c$th place which is colored either by color 1 or
color 2, is actually colored by 1, and it is the $a$th place which is colored by 1.
Noting that $\widetilde{{\bm I}_{k_1}^{(1)}}$ is the set induced by the map
\eqref{maptoinduce}, we find that $i_c^{(2)}=i_a^{(1)}$
corresponds to $c=\widetilde{I_a^{(1)}}$.
By similar considerations, we note the following correspondence
\begin{align}
i_c^{(2)}=i_a^{(1)} \longleftrightarrow c=\widetilde{I_a^{(1)}}, 
\ \ \
i_c^{(2)}>i_a^{(1)} \longleftrightarrow c>\widetilde{I_a^{(1)}},
\ \ \
i_c^{(2)}<i_a^{(1)} \longleftrightarrow c<\widetilde{I_a^{(1)}}, 
\label{useformatch3} \\
i_c^{(3)}=i_a^{(2)} \longleftrightarrow c=I_a^{(2)},
\ \ \
i_c^{(3)}>i_a^{(2)} \longleftrightarrow c>I_a^{(2)},
\ \ \
i_c^{(3)}<i_a^{(2)} \longleftrightarrow c<I_a^{(2)}.
\label{useformatch4}
\end{align}
We also note that $j(I,1,a)$ is always 1 and
$j(I,2,a)$ in \cite{RTV2} corresponds to $i_{I_a^{(2)}}^{(2)}$ in this paper.

From the correspondences
\eqref{useformatch1}, \eqref{useformatch2}, \eqref{useformatch3}, \eqref{useformatch4},
we can rewrite the case $N=3$ of the elliptic weight functions
\eqref{weightforcomparison} as
\begin{align}
&\overline{W}_{I}^{\mathrm{ell}}({\bm t},{\bm z},h,{\bm \mu}) \nonumber \\
=&\mathrm{Sym}_{t^{(1)}} \mathrm{Sym}_{t^{(2)}}
\prod_{a=1}^{\lambda^{(1)}}
\Bigg(
\Bigg(
\prod_{c=1}^{\widetilde{I_a^{(1)}}-1}
\vartheta(ht_c^{(2)}/t_a^{(1)})
\Bigg)
\vartheta(
t_{\widetilde{I_a^{(1)}}}^{(2)}/t_a^{(1)} h^{2a-\widetilde{I_a^{(1)}}}
\mu_2/\mu_1) \nonumber \\
&\times \Bigg(
\prod_{c=\widetilde{I_a^{(1)}}+1}^{\lambda^{(2)}} \theta(t_c^{(2)}/t_a^{(1)})
\Bigg)
\Bigg(
\prod_{b=a+1}^{\lambda^{(1)}} \frac{\vartheta(ht_b^{(1)}/t_a^{(1)})}
{\vartheta(t_b^{(1)}/t_a^{(1)})} \Bigg) \Bigg) \nonumber \\
&\times \prod_{a=1}^{\lambda^{(2)}}
\Bigg(
\Bigg(
\prod_{c=1}^{I_a^{(2)}-1}
\vartheta(ht_c^{(3)}/t_a^{(2)})
\Bigg)
\vartheta(
t_{I_a^{(2)}}^{(3)}/t_a^{(2)} h^{
a-I_a^{(2)}+c(I_a^{(2)},i_{I_a^{(2)}}^{(2)})
}
\mu_3/\mu_{i_{I_a^{(2)}}^{(2)}}) \nonumber \\
&\times \Bigg(
\prod_{c=I_a^{(2)}+1}^{\lambda^{(3)}} \theta(t_c^{(3)}/t_a^{(2)})
\Bigg)
\Bigg(
\prod_{b=a+1}^{\lambda^{(2)}} \frac{\vartheta(ht_b^{(2)}/t_a^{(2)})}
{\vartheta(t_b^{(2)}/t_a^{(2)})} \Bigg) \Bigg) \nonumber \\
=&\sum_{\sigma_1 \in S_{\lambda^{(1)}}}
\sum_{\sigma_2 \in S_{\lambda^{(2)}}}
\prod_{a=1}^{\lambda^{(1)}}
\Bigg(
\Bigg(
\prod_{c=1}^{\widetilde{I_a^{(1)}}-1}
\vartheta(ht_{\sigma_2(c)}^{(2)}/t_{\sigma_1(a)}^{(1)})
\Bigg)
\vartheta(
t_{\sigma_2(\widetilde{I_a^{(1)}})}^{(2)}/t_{\sigma_1(a)}^{(1)}
h^{2a-\widetilde{I_a^{(1)}}}
\mu_2/\mu_1) \nonumber \\
&\times \Bigg(
\prod_{c=\widetilde{I_a^{(1)}}+1}^{\lambda^{(2)}} \theta(t_{\sigma_2(c)}^{(2)}/t_{\sigma_1(a)}^{(1)})
\Bigg)
\Bigg(
\prod_{b=a+1}^{\lambda^{(1)}} \frac{\vartheta(ht_{\sigma_1(b)}^{(1)}/t_{\sigma_2(a)}^{(1)})}
{\vartheta(t_{\sigma_1(b)}^{(1)}/t_{\sigma_1(a)}^{(1)})} \Bigg) \Bigg) \nonumber \\
&\times \prod_{a=1}^{\lambda^{(2)}}
\Bigg(
\Bigg(
\prod_{c=1}^{I_a^{(2)}-1}
\vartheta(ht_c^{(3)}/t_{\sigma_2(a)}^{(2)})
\Bigg)
\vartheta(
t_{I_a^{(2)}}^{(3)}/t_{\sigma_2(a)}^{(2)} h^{
a-I_a^{(2)}+c(I_a^{(2)},i_{I_a^{(2)}}^{(2)})
}
\mu_3/\mu_{i_{I_a^{(2)}}^{(2)}}) \nonumber \\
&\times \Bigg(
\prod_{c=I_a^{(2)}+1}^{\lambda^{(3)}} \theta(t_c^{(3)}/t_{\sigma_2(a)}^{(2)})
\Bigg)
\Bigg(
\prod_{b=a+1}^{\lambda^{(2)}} \frac{\vartheta(ht_{\sigma_2(b)}^{(2)}/t_{\sigma_2(a)}^{(2)})}
{\vartheta(t_{\sigma_2(b)}^{(2)}/t_{\sigma_2(a)}^{(2)})} \Bigg) \Bigg).
\label{comparisonlast}
\end{align}

To compare \eqref{comparisonlast}
with the elliptic multivariable functions
comining from the Foda-Manabe type partition functions \eqref{forcomparison},
we need to 
change $\lambda^{(1)}$,
$\lambda^{(2)}$ and $\lambda^{(3)}$
to $k_1$, $k_2$ and $L_2$ respectively
by the correspondence \eqref{preuseformatch2}.
Finally, we
switch from the multiplicative notation
to the additive notation.
Making the following change of variables
$t_a^{(1)}=e^{-2 \pi i z_a^{(1)}}$,
$t_a^{(2)}=e^{-2 \pi i z_a^{(2)}}$,
$t_a^{(3)}=e^{-2 \pi i w_a^{(2)}}$,
$h=e^{2 \pi i \gamma}$,
$\mu_1=e^{2 \pi i \lambda_1}$,
$\mu_2=e^{2 \pi i \lambda_2}$,
$\mu_3=e^{2 \pi i \lambda_3}$,
and switching to the additive notation,
we find that \eqref{comparisonlast}
becomes \eqref{forcomparison}.

\section{Conclusion}
In this paper, we introduced and analyzed partition functions associated with $E_{\tau,\eta}(gl_3)$
which is an elliptic analogue 
of the one recently introduced by Foda and Manabe \cite{FM}.
For the analysis, we developed a nested version of the Izergin-Korepin method
\cite{Ko,Iz}
which is a higher rank extension of the method for the wavefunctions of six-vertex type models \cite{Motegi,Motegi2}.
The partition functions are explicitly expressed
as symmetrization of elliptic multivariable functions
over two sets of variables.
Multivariable functions which have multiple sets of symmetric variables appear
as explicit representations for partition functions of Foda-Manabe type
\cite{FM}.
In the context of quantum integrable models,
symmteric functions which have multiple sets of symmetric variables
also appear in the trigonometric weight functions and elliptic weight functions
\cite{TarVarSigma,Konno1,Konno2,FRV,RTV2,TV,RTV1,Sh},
which originally appeared in the integral representation of
solutions to the $q$-KZ equation, and also become the elliptic stable envelopes
for the cotangent bundles of flag varieties.
Elliptic stable envelopes are certain elliptic classes proposed in \cite{AO}
to study an elliptic extension of a program initiated in
\cite{MO} to relate quantum torus equivariant cohomology
of quiver varieties and representation theory of quantum groups,
which is
regarded as a mathematical formulation of the Bethe/Gauge correspondence
\cite{NS1,NS2}.
The elliptic weight functions was introduced as an elliptic analogue
of the trigonometric weight functions, and it was shown that
the trigonometric weight functions have connections with the
off-shell Bethe wavefunctions of trigonometric integrable models \cite{TarVarSigma}.
From this fact, it is supposed that the case $L_1=0$
of the elliptic multivariable functions \eqref{multivariablefunction}
have connections with the elliptic weight functions,
since the off-shell Bethe wavefunctions of the trigonometric model
correspond to the case $L_1=0$ of the partition functions
of Foda-Manabe type.
We showed that this special case corresponds
to the elliptic weight functions introduced in
the studies by Rim\'anyi-Tarasov-Varchenko, Konno, Felder-Rim\'anyi-Varchenko
\cite{Konno1,Konno2,FRV,RTV2},
hence the partition functions introduced in this paper
serve as representations of objects in geometric representation theory.

\section*{Acknowledgements}
This work was partially supported by Grant-in-Aid for Scientific Research
(C) No. 18K03205 and No. 16K05468.


\begin{thebibliography}{00}
%

\bibitem{Bethe}
H. Bethe,
Z. Phys. {\bf 71}, 205 (1931).
%
\bibitem{Baxter}
R.J. Baxter,
{\it Exactly Solved Models in Statistical Mechanics}
(Academic Press, London, 1982).
%
\bibitem{KBI}
V.E. Korepin, N.M. Bogoliubov and A.G. Izergin
{\it Quantum Inverse Scattering Method and Correlation functions}
(Cambridge University Press, Cambridge, 1993).
%
\bibitem{Bogo}
N.M. Bogoliubov,
J. Phys. A: Math. Gen. {\bf 38}, 9415 (2005).
%
\bibitem{ShigechiUchiyama}
K. Shigechi, M. Uchiyama,
J. Phys. A: Math. Gen. {\bf 38}, 10287 (2005).
%
\bibitem{BeWh}
D. Betea and M. Wheeler,
J. Comb. Th. Ser. A {\bf 137}, 126 (2016).
%
\bibitem{BWZ}
D. Betea, M. Wheeler and P. Zinn-Justin,
J. Alg. Comb. {\bf 42}, 555 (2015).
%
\bibitem{WZnew}
M. Wheeler and P. Zinn-Justin,
Adv. Math. {\bf 299}, 543 (2016).
%
\bibitem{vDE}
J.F. van Diejen and E. Emsiz,
Commun. Math. Phys. {\bf 350}, 1017 (2017).
%
\bibitem{MS}
K. Motegi and K. Sakai,
J. Phys. A: Math. Theor. {\bf 46}, 355201 (2013).
%
\bibitem{MS2}
K. Motegi and K. Sakai,
J. Phys. A: Math. Theor. {\bf 47}, 445202 (2014).
%
\bibitem{Korff}
C. Korff,
Lett. Math. Phys. {\bf 104}, 771 (2014).
%
\bibitem{GK2}
V. Gorbounov and C. Korff,
Adv. Math. {\bf 313}, 282 (2017).
%
\bibitem{Borodin}
A. Borodin,
Adv. in Math. {\bf 306}, 973 (2017).
%
\bibitem{Borodin2}
A. Borodin,
Symmetric elliptic functions, IRF models, and dynamic exclusion processes,
e-print arXiv:1701.05239.
%
\bibitem{BP1}
A. Borodin and L. Petrov
Sel. Math. New Ser. {\bf 24} 751 (2016).
%
\bibitem{BBF}
B. Brubaker, D. Bump and S. Friedberg,
{\it Commun. Math. Phys.} {\bf 308}, 281 (2011).
%
\bibitem{Iv}
D. Ivanov,
{\it Symplectic Ice.} in:
Multiple Dirichlet series, $L$-functions
and automorphic forms, vol 300 of Progr. Math. 
(Birkh\"auser/Springer,
New York, 2012) pp. 205-222.
%
\bibitem{BBCG}
B. Brubaker, D. Bump, G. Chinta and P.E. Gunnells P E,
{\it Metaplectic Whittaker Functions and Crystals of Type B.} in:
Multiple Dirichlet series, $L$-functions
and automorphic forms, vol 300 of Progr. Math. 
(Birkh\"auser/Springer,
New York, 2012) pp. 93-118.
%
\bibitem{Tabony}
S.J. Tabony
{\it Deformations of characters, metaplectic Whittaker functions
and the Yang-Baxter equation,} PhD. Thesis,
Massachusetts Institute of Technology, USA 2011.
%
\bibitem{BMN}
D. Bump, P. McNamara and M. Nakasuji,
Comm. Math. Univ. St. Pauli {\bf 63}, 23 (2014).
%
\bibitem{TarVarSigma}
V. Tarasov and A. Varchenko,
SIGMA {\bf 9}, 048 (2013).
%
\bibitem{FWZ}
O. Foda, M. Wheeler, and M. Zuparic,
J. Stat. Mech.: Theory Exp. 2008, P02001.
%
\bibitem{FW}
O. Foda and M. Wheeler,
Nucl. Phys. B {\bf 871}, 330 (2013).
%
\bibitem{WZ}
M. Wheeler and P. Zinn-Justin,
J. Reine Angew. Math. {\bf 757}, 159 (2019).
%
\bibitem{Tak}
Y. Takeyama,
Funkcialaj Ekvacioj, {\bf 61}, 349 (2018).
%
\bibitem{BW}
A. Borodin and M. Wheeler,
Coloured stochastic vertex models and their spectral theory,
e-print arXiv:1808.01866.
%
\bibitem{metaplectic}
B. Brubaker, V. Buciumas, D. Bump and N. Gray,
Comm. Numb. Theor. Phys.
{\bf 13}, 101 (2019).
%
\bibitem{BBBG}
B. Brubaker, V. Buciumas, D. Bump and H.P.A. Gustafsson,
Colored five-vertex models and Demazure atoms,
e-print arXiv:1902.01795.
%
\bibitem{BBBG2}
B. Brubaker, V. Buciumas, D. Bump and H.P.A. Gustafsson,
Colored Vertex Models and Iwahori Whittaker Functions,
e-print arXiv:1906.04140.
%
\bibitem{BSW}
V. Buciumas, T. Scrimshaw, K. Weber, Colored vertex models
and Lascoux polynomials and atoms, e-print arXiv:1908.07364.
%
\bibitem{FM}
O. Foda and M. Manabe,
J. High Energ. Phys. {\bf 2019}, 36 (2019).
%
\bibitem{NS1}
N. Nekrasov and S. Shatashvili,
Nucl. Phys. Proc. Supp. {\bf 192-193}, 91 (2009).
%
\bibitem{NS2}
N. Nekrasov and S. Shatashvili,
Prog. Theor. Phys. Supp. {\bf 177}, 105 (2009).
%
\bibitem{Felder}
G. Felder, Elliptic quantum groups. In: Iagolnitzer, D. (ed.)
Proceedings of the ICMP, Paris 1994, pp. 211-218. Intern. Press, Cambridge, MA (1995)
%
\bibitem{FV1}
G. Felder and A. Varchenko,
Comm. Math. Phys. {\bf 181}, 741 (1996).
%
\bibitem{FV2}
G. Felder and A. Varchenko,
Nucl. Phys. B, {\bf 480}, 485 (1996).
%
\bibitem{Ca}
A. Cavalli,
On representations of the Elliptic Quantum Group $E_{\gamma,\tau}(gl_N)$,
PhD thesis, 2001, ETH Z\"urich.
%
\bibitem{Ko}
V.E. Korepin,
Commun. Math. Phys. {\bf 86}, 391 (1982).
%
\bibitem{Iz}
A. Izergin
Sov. Phys. Dokl. {\bf 32}, 878 (1987).
%
\bibitem{Br}
D. Bressoud,
{\it Proofs and confirmations:
The story of the alternating sign matrix conjecture}
(MAA Spectrum,
Mathematical Association of America,
Washington, DC, 1999).
%
\bibitem{Ku1}
G. Kuperberg,
Int. Math. Res. Not. {\bf 3}, 123 (1996).
%
\bibitem{Ku2}
G. Kuperberg,
Ann. Math. {\bf 156}, 835 (2002).
%
\bibitem{Okada}
S. Okada,
J. Alg. Comb. {\bf 23}, 43 (2001).
%
\bibitem{CP}
F. Colomo and A.G. Pronko,
J. Stat. Mech.:Theor. Exp. P01005 (2005).
%
\bibitem{KZ}
V. Korepin and P. Zinn-Justin,
J. Phys. A {\bf 33}, 7053 (2000).
%
\bibitem{BL}
P. Bleher and K. Liechty
J. Stat. Phys. {\bf 134}, 463 (2009).
%
\bibitem{HK1}
A. Hamel and R.C. King,
J. Algebraic Comb. {\bf 16}, 269 (2002).
%
\bibitem{HK2}
A. Hamel and R.C. King
J. Algebraic Comb. {\bf 21}, 395 (2005).
%
\bibitem{Tsuchiya}
O. Tsuchiya,
J. Math. Phys. {\bf 39}, 5946 (1998).
%
\bibitem{Wheeler}
M. Wheeler,
Nucl.Phys.B {\bf 852}, 468 (2011).
%
\bibitem{PRS}
S. Pakuliak, V. Rubtsov and A. Silantyev,
J. Phys. A:Math. Theor. {\bf 41}, 295204 (2008).
%
\bibitem{Ros}
H. Rosengren,
Adv. Appl. Math. {\bf 43}, 137 (2009).
%
\bibitem{FK}
F. Filali and N. Kitanine,
J. Stat. Mech. L06001 (2010).
%
\bibitem{Chinesegroup}
W-L. Yang, X. Chen, J. Feng, K. Hao, K-J. Shi, C-Y. Sun, Z-Y. Yang
and Y-Z. Zhang,
Nucl. Phys. B {\bf 847}, 367 (2011).
%
\bibitem{Chinesegroup2}
W-L. Yang, X. Chen, J. Feng, K. Hao, K. Wu, Z-Y. Yang
and Y-Z. Zhang,
Nucl. Phys. B {\bf 848}, 523 (2011).
%
\bibitem{Galleasone}
W.Galleas,
Nucl. Phys. B {\bf 858}, 117 (2012).
%
\bibitem{Galleasthree}
W. Galleas,
Phys. Rev. E {\bf 94}, 010102(R) (2016).
%
\bibitem{GL}
W. Galleas, J. Lamers, Nucl. Phys. B {\bf 886}, 1003 (2014).
%
\bibitem{Lamers}
J. Lamers, Nucl. Phys. B {\bf 901}, 556 (2015).
%
\bibitem{Motegi}
K. Motegi,
J. Math. Phys. {\bf 59}, 053505 (2018).
%
\bibitem{Motegi2}
K. Motegi,
Prog. Theor. Exp. Phys. {\bf 2017}, 123A01 (2017).
%
\bibitem{Konno1}
H. Konno,
J. Int. Syst. {\bf 2}, xyx011 (2017).
%
\bibitem{Konno2}
H. Konno,
J. Int. Syst. {\bf 3}, xyy012 (2018).
%
\bibitem{FRV}
G.  Felder, R. Rim\'anyi and A. Varchenko,
SIGMA {\bf 14}, 41 (2018).
%
\bibitem{RTV2}
R. Rim\'anyi, V. Tarasov and A. Varchenko,
Sel. Math. {\bf 25}, 16  (2019).
%
\bibitem{AO}
M. Aganagic and A. Okounkov,
Elliptic stable envelopes,
e-print arXiv:1604.00423.
%
\bibitem{MO}
D. Maulik and A. Okounkov,
Quantum groups and quantum cohomology,
Ast\'erisque, {\bf 408} (2019).
%
\bibitem{FSfelderhof}
G. Felder and A. Schorr,
J. Phys. A: Math.Gen. {\bf 32}, 8001 (1999).
%
\bibitem{ABF}
G.E. Andrews, R.J. Baxter and P.J. Forrester,
J. Stat. Phys. {\bf 35}, 193 (1984).
%
\bibitem{eightvertex}
R.J. Baxter,
Ann. Phys. {\bf 70}, 193 (1972).
%
\bibitem{FIJKMY}
O. Foda, K. Iohara, M. Jimbo, R. Kedem, T. Miwa and H. Yan,
Lett. Math. Phys. {\bf 32}, 259 (1994).
%
\bibitem{Fr}
C. Fronsdal,
Lett. Math. Phys. {\bf 40}, 117 (1997).
%
\bibitem{Konno}
H. Konno,
Comm. Math. Phys. {\bf 195}, 373 (1998).
%
\bibitem{JKOS}
M. Jimbo, H. Konno, S. Odake and J. Shiraishi, 
Trans. Groups. {\bf 4}, 303 (1999). 
%
\bibitem{DJKMO}
E. Data, M. Jimbo, A. Kuniba, T. Miwa and M. Okado,
Nucl. Phys.B {\bf 290}, 231 (1987).
%
\bibitem{JKMO}
M. Jimbo, A. Kuniba, T. Miwa and M. Okado,
Comm. Math. Phys. {\bf 119}, 543 (1988).
%
\bibitem{TV}
V. Tarasov and A. Varchenko,
Leningrad Math. J. {\bf 6}, 275 (1994).
%
\bibitem{Reshetikhin}
N. Yu. Reshetikhin,
J. Sov. Math. {\bf 46}, 1694 (1989).
%
\bibitem{RTV1}
R. Rim\'anyi, V. Tarasov and A. Varchenko,
J. Geom. Phys. {\bf 94}, 81 (2015).
%
\bibitem{Sh}
D. Shenfeld,
Abelianization of Stable Envelopes in Symplectic Resolutions, PhD thesis, Princeton, 2013.
%







\end{thebibliography}
\end{document}